\documentclass{article}

\usepackage{microtype}
\usepackage{graphicx}
\usepackage{subcaption}
\usepackage{booktabs} 

\usepackage{hyperref}
\usepackage{colortbl}
\usepackage{xcolor}
\definecolor{headerblue}{RGB}{0,112,192}
\usepackage{array}

\usepackage[accepted]{icml2026}
\usepackage{booktabs}
\usepackage{tabularx}
\usepackage{makecell}

\usepackage{amsmath}
\usepackage{amssymb}
\usepackage{mathtools}
\usepackage{amsthm}
\usepackage{multirow}
\usepackage{url}
\usepackage[capitalize,noabbrev]{cleveref}

\theoremstyle{plain}

\theoremstyle{definition}

\theoremstyle{remark}

\usepackage[most]{tcolorbox}
\tcbset{
  mybox/.style={
    enhanced,
    breakable,
    colback=blue!4,             
    colframe=blue!65!black,     
    colbacktitle=blue!70!black,
    coltitle=white,            
    fonttitle=\bfseries,       
    boxrule=0.8pt,
    arc=2mm,
    left=2mm, right=2mm, top=1mm, bottom=1mm,
    title style={font=\bfseries},
  },
    redbox/.style={
    enhanced,
    breakable,
    colback=red!5,
    colframe=red!70!black,
    colbacktitle=red!75!black,
    coltitle=white,
    fonttitle=\bfseries,
    boxrule=0.8pt,
    arc=2mm,
    left=2mm, right=2mm, top=1mm, bottom=1mm,
    title style={font=\bfseries},
  }
}

\usepackage[textsize=tiny]{todonotes}

\icmltitlerunning{StitchCUDA: An Automated End-to-End Machine Learning Program Generation Framework with Rubric-based Agentic Reinforcement Learning}

\begin{document}
\twocolumn[
  \icmltitle{StitchCUDA: An Automated Multi-Agents End-to-End GPU Programing Framework with Rubric-based Agentic Reinforcement Learning}

  \icmlsetsymbol{equal}{*}
  \begin{icmlauthorlist}
    \icmlauthor{Shiyang Li$^*$}{umn}
    \icmlauthor{Zijian Zhang$^*$}{umn}
    \icmlauthor{Winson Chen}{umn}
    \icmlauthor{Yuebo Luo}{umn}
    \icmlauthor{Mingyi Hong}{umn}
    \icmlauthor{Caiwen Ding}{umn}
  \end{icmlauthorlist}
  \icmlaffiliation{umn}{University of Minnesota-Twin Cities, Minnesota, USA.  {*}means equal contribution}

  \icmlcorrespondingauthor{Shiyang Li}{li004074@umn.edu}
  \icmlcorrespondingauthor{Zijian Zhang}{zha00175@umn.edu}
  \icmlcorrespondingauthor{Winson Chen}{chen9619@umn.edu}
  \icmlcorrespondingauthor{Yuebo Luo}{luo00466@umn.edu}
  \icmlcorrespondingauthor{Mingyi Hong}{mhong@umn.edu}
  \icmlcorrespondingauthor{Caiwen Ding}{dingc@umn.edu}

  \vskip 0.1in
]


\printAffiliationsAndNotice{}  

\newcommand{\blue}{\color{blue}}
\newcommand{\red}{\color{red}}

\begin{abstract}

Modern machine learning (ML) workloads increasingly rely on GPUs, yet achieving high end-to-end performance remains challenging due to dependencies on both GPU kernel efficiency and host-side settings.
Although LLM-based methods show promise on automated GPU kernel generation, prior works mainly focus on single-kernel optimization and do not extend to end-to-end programs, hindering practical deployment.

To address the challenge, in this work, we propose \textsc{StitchCUDA}, a multi-agent framework for end-to-end GPU program generation, with three specialized agents: a \textit{Planner} to orchestrate whole system design, a \textit{Coder} dedicated to implementing it step-by-step, and a \textit{Verifier} for correctness check and performance profiling using Nsys/NCU. To fundamentally improve the \textit{Coder}'s ability in end-to-end GPU programming, \textsc{StitchCUDA} integrates rubric-based agentic reinforcement learning over two atomic skills, task-to-code generation and feedback-driven code optimization, with combined rubric reward and rule-based reward from real executions. Therefore, the \textit{Coder} learns how to implement advanced CUDA programming techniques (e.g., custom kernel fusion, cublas epilogue), and we also effectively prevent \textit{Coder}'s reward hacking (e.g., just copy PyTorch code or hardcoding output) during benchmarking.
Experiments on KernelBench show that \textsc{StitchCUDA} achieves nearly 100\% success rate on end-to-end GPU programming tasks, with 1.72$\times$  better speedup over the multi-agent baseline and 2.73$\times$ than the RL model baselines.


\end{abstract}

\section{Introduction}

Modern machine learning (ML) workloads increasingly rely on GPUs, yet developing high-performance GPU programs remains challenging, especially for end-to-end workloads where performance depends on both GPU kernel efficiency and host-side settings (e.g., memory allocation and CPU–GPU overlap). 

While LLM-based code generation offers a promising path to automate GPU programming, existing methods—including multi-agent systems~\cite{cudaforge,cupilot,astra} and domain-specific fine-tuning or RL~\cite{kevin,cudal2}—focus on single-kernel optimization, limiting them to Level 1/2 tasks in KernelBench~\cite{kernelbench} (e.g., a 3D max-pooling kernel).

However, the central challenge lies in moving beyond kernel-level generation to full end-to-end GPU programs, which these methods do not support. In contrast, KernelBench~\cite{kernelbench} Level 3/4 workloads involve multiple interacting kernels and impose fundamentally different system-level requirements.
For example, the entire VisionTransformer model architecture~\cite{vit} in level 3,  whose end-to-end performance is dominated by factors that go beyond any single kernel: kernel fusion's boundaries, launch config, CPU–GPU synchronization, and data movement, which has been proven critical for GPU program performance~\cite{emogi,mobius,native,liberator}.

\begin{figure}[t]
    \centering
    \includegraphics[width=\linewidth]{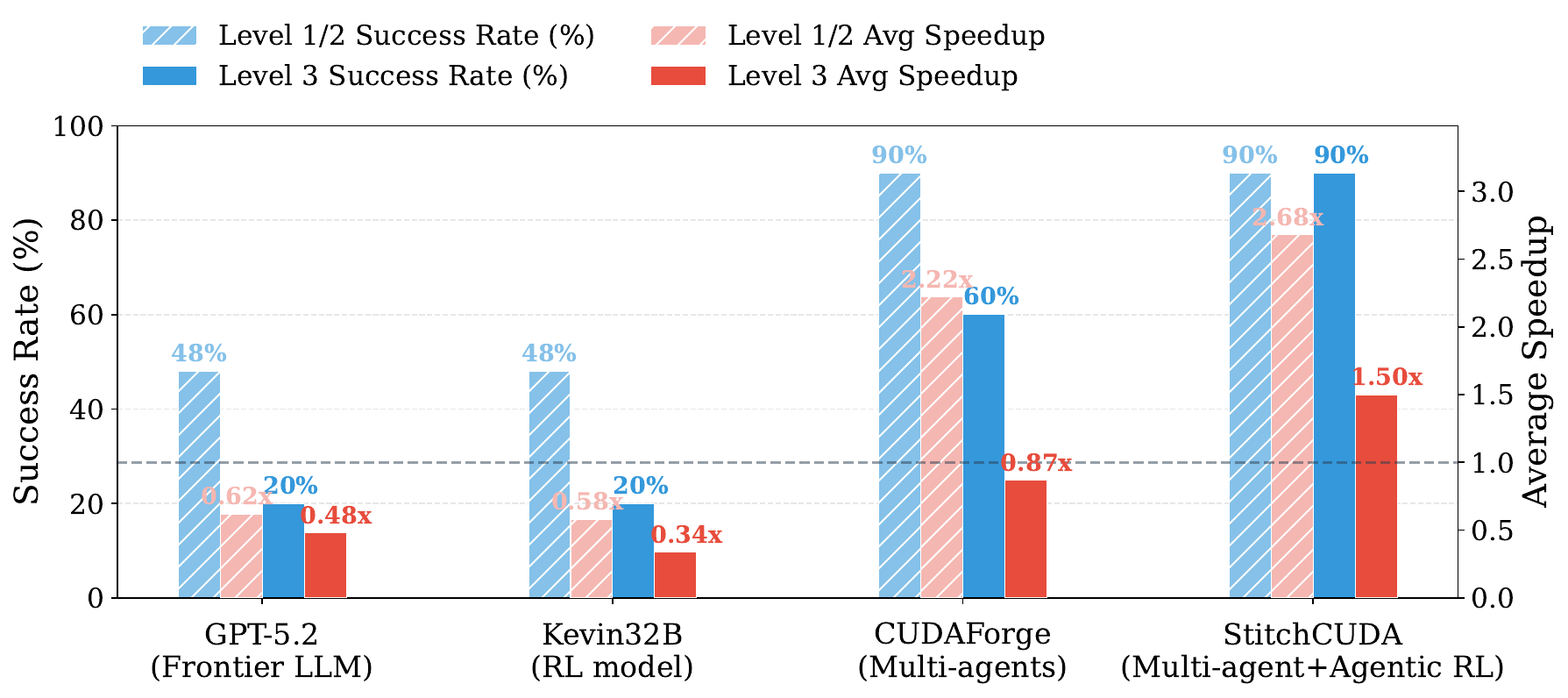}
    \caption{The performance of various automated GPU program generation methods on KernelBench Level 1/2/3. Achieved average speedup is relative to PyTorch eager mode code, using NVIDIA H200 GPU.}
    \label{fig:l3}
    \vspace{-18pt}
\end{figure}

Recent multi-agent approaches such as CUDAForge and QiMeng~\cite{cudaforge,qimeng} leverage fine-grained prompt engineering and task decomposition (e.g., Coder and Verifier agents), showing promise on single kernels. However, they lack mechanisms for enforcing cross-kernel optimization and host-side orchestration. As shown in Fig.~\ref{fig:l3}, CUDAForge achieves near-perfect correctness with strong speedups on KernelBench Level 1/2, but struggles on Level 3 with lower success rates and performance gains.

In addition, according to our observation discussed in Section~\ref{sec:discuss}, the Coder's ability also limits the overall performance of current methods. 
A common approach to improve model
ability in a specific domain is Reinforcement Learning (RL), especially when rewards are verifiable
(a.k.a RLVR). Prior work~\cite{kevin,cudal2,li2025cudal1improvingcudaoptimization}
uses functional correctness and speedup over reference implementations as
verifiable rewards to train LLMs with GRPO, yielding measurable gains. However,
these rule-based rewards are vulnerable to \emph{reward hacking}, leading to undesirable behaviors such as writing PyTorch-only code or hardcoding the output without implementing a real program (see Appendix~\ref{app:rewardhack} for examples). Moreover, they
can induce degenerate solutions (e.g., only replacing a trivial operator such as a standalone \texttt{ReLU} in a Neural Network) that obtain nontrivial
reward yet do not correspond to meaningful improvements in kernel quality. Thus, an RL-trained model could achieve a high success rate, but fails to deliver effective optimization, as Kevin-32B shown in Fig.\ref{fig:l3}. 

More fundamentally, existing RLVR formulations mainly focus on kernel generation alone, without learning to incorporate \emph{structured} feedback (e.g., compiler diagnostics and profiling bottlenecks) to 
optimize code accordingly. As a result, the model often fails to reliably follow concrete feedback and implement targeted optimizations in an agentic framework, which directly undermines end-to-end optimization performance.
Agentic RL is a promising alternative because it directly trains the model to
operate within an agentic framework, where the model interacts with the environment, gets feedback, and conducts the next action. However, collecting multi-turn rollouts incurs
substantial computational overhead, making training very inefficient.

We summarized the following challenges according to our experiments and observations:

    \noindent \textbf{(C1) End-to-end program requires global coordination.}
    Unlike single-kernel optimization, end-to-end GPU program performance is dominated by intra-kernel decisions (e.g., kernel fusion boundaries, and memory footprint among kernels) and host-side orchestration; the automated generation needs to reason over a \emph{program-level} state rather than dealing with each kernel design individually. This complex end-to-end system design cannot be well addressed by one-shot LLM inference or a simple self-refine loop.
.
    
    \noindent \textbf{(C2) Coder's CUDA-specific coding capability needs to be improved beyond prompting.} 
    Multi-agent decomposition can surface feedback from other agents to guide Coder, but without parameter updates, the Coder often cannot reliably execute nontrivial CUDA transformations (e.g., deriving a correct tiling strategy from profiling hints), thereby becoming the primary bottleneck in practice.

    \noindent \textbf{(C3) Direct RLVR fails to improve Coder's capability in end-to-end setting.} 
    Existing RLVR approaches for CUDA code generation are vulnerable to reward hacking and can induce degenerate solutions. More fundamentally, coders are not explicitly trained to interpret \emph{structured} execution feedback and to apply targeted improvements. While agentic RL could in principle address this gap, multi-turn rollouts are, however, prohibitively costly in realistic CUDA environments, making training extremely inefficient.

To apply LLMs in automated end-to-end GPU program generation and optimization, we propose StitchCUDA, a multi-agent framework integrated with rubric-based agentic reinforcement learning. StitchCUDA instantiates three specialized agents: a \textit{Planner} that decomposes the required task in reference PyTorch code into a program specification (kernel fusion boundaries, tensor shapes/layout contracts, and CPU-GPU overlapping), a \textit{Coder} that implements the host code and GPU kernels accordingly, and a \textit{Verifier} that enforces correctness checks and do the performance analysis using Nsight Systems (Nsys) and Nsight Compute (NCU), enabling an iterative “plan–code–profile–refine” loop for end-to-end GPU programming. 

Beyond multi-agent orchestration, we further improve the \textit{Coder} via rubric-based agentic reinforcement learning, enabling end-to-end optimization on a sequence of code-generation and optimization actions. To alleviate rollout overhead, we decompose multi-turn agentic RL into two \emph{atomic skills}: \textbf{(1) from-scratch generation}, translating high-level GPU programming tasks and reference code into a correct CUDA implementation; and \textbf{(2) feedback-driven optimization}, incorporating structured execution feedback (e.g., compiler diagnostics and profiling bottlenecks) to fix bugs and improve performance. We collect single-turn training data for both skills during workflow execution to train \textit{Coder}, which is substantially more efficient than multi-turn agentic RL.
Further, to address reward hacking and degenerate behaviors,
we combine rule-based rewards from real executions (functional correctness/measured end-to-end speedup) with expert-aligned rubric rewards produced by an advanced LLM (details in Section~\ref{RLdesign}), avoiding reward hacking and encouraging end-to-end system optimization.

Our key contributions are summarized below:
\begin{itemize}

\vspace{-0.2cm}
    \item We propose StitchCUDA, a multi-agent system that generates complete end-to-end GPU programs from task requirements and PyTorch references, integrated with an agentic RL technique to effectively improve the Coder's capability in this domain. We decompose end-to-end GPU programming tasks into three agents. We also prepare expert-level chains of thoughts (CoT) in prompt engineering and effective expertise tools for them, to better guide the Coder to address end-to-end program bottlenecks.   

    \vspace{-0.2cm}
    \item We decompose multi-turn agentic RL into atomic skills, substantially reducing rollout overhead while better aligning with the end-to-end task. Combined with fine-grained rubric rewards, our approach mitigates reward hacking/degenerate behavior and performs better.

    \vspace{-0.2cm}
    \item Comprehensive experiment results show that StitchCUDA performs outstandingly in the end-to-end GPU programming task. In our test sets derived from KernelBench Level 3, StitchCUDA achieves nearly 100\% success rate, and 1.5$\times$ average speedup than PyTorch eager (1.72$\times$$\uparrow$ than multi-agent approach, and 2.73$\times$$\uparrow$ than RL model).  
\end{itemize}
Code of the \textsc{StitchCUDA} framework is avalaible at https://github.com/UMN-APEX-Lab/StitchCUDA.

\section{Background}

\label{sec:background}

\textbf{Benchmarks and task formulations.}
Recent work has standardized the evaluation of LLMs for CUDA programming via \emph{KernelBench}~\cite{kernelbench}, which frames automated kernel generation as translating a PyTorch reference into a CUDA extension and evaluates its functional correctness and speed relative to the reference. KernelBench Level~1/2 primarily emphasizes \emph{single-kernel} generation and a few fusion opportunities under fixed interfaces and micro-benchmark objectives, making the optimization target comparatively local and well-scoped. In contrast, Level~3/4 increasingly resemble program synthesis rather than kernel synthesis, where multiple kernels and host orchestration jointly determine end-to-end behavior and performance. 

\textbf{LLM Agents and workflow.}
A common approach is to externalize the kernel-development loop into an agentic workflow. CUDAForge \cite{cudaforge} exemplifies this design for kernel optimization: specialized agents (a coding agent and a feedback agent) iteratively refine an existing CUDA kernel, using correctness checks and profiling feedback to guide improvement. Moreover,
cuPilot \cite{cupilot}, astra \cite{astra}, and QiMeng \cite{qimeng} further push agentic optimization toward evolutionary search and roofline-guided prompting.
These systems demonstrate that decomposition and tool-augmented iteration can improve success rates and speedups in practice, but they largely operate within a single-kernel scope. 

However, multi-agent approaches face several structural limitations scaling to end-to-end GPU programs. These workflows are prone to \emph{local optima} and \emph{specification drift}: iterative edits may overfit to specific inputs and profiling artifacts. Besides, as the search space grows (multiple kernels, host orchestration, and library calls), the coding agent suffers from long code context and tool traces, increasing the likelihood of inconsistent implementation, redundant edits, and unstable convergence, collectively limiting scalability beyond single-kernel optimization.

\textbf{RL-training paradigms.}
A growing line of work~\cite{kevin,cudal2,li2025cudal1improvingcudaoptimization} improves LLM-based kernel generators via post-training on \emph{verifiable} execution signals. Kevin~\cite{kevin} proposes a multi-turn RL recipe tailored to kernel generation, where reward is defined by functional correctness and runtime speedup over a reference implementation. Similarly, CUDA-L1~\cite{li2025cudal1improvingcudaoptimization} and CUDA-L2~\cite{cudal2} adopt rule-based rewards for GRPO training, targeting general kernel optimization and HGEMM, respectively.

Despite promising results, RL-based methods face three recurring limitations: (i) correctness/speedup rewards are prone to reward hacking and degenerate behavior; (ii) coders are typically not trained to interpret structured execution feedback, limiting their effectiveness in end-to-end optimization; and (iii) although agentic RL could address this gap in principle, multi-turn rollouts in realistic CUDA environments are prohibitively costly, making training inefficient.

 \begin{figure*}[ht]
    \centering
    \includegraphics[width=\linewidth]{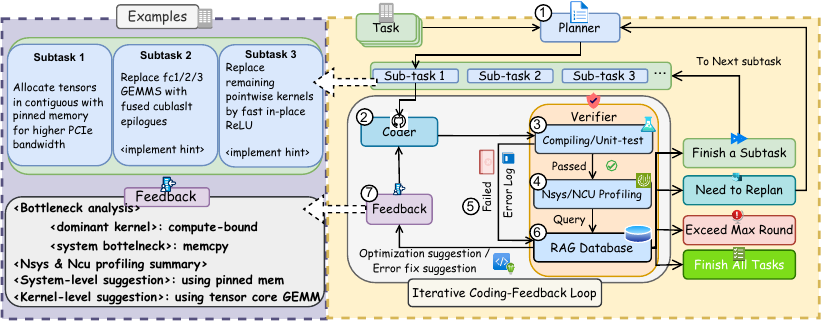}
    \caption{StitchCUDA: \textcircled{1} The Planner profiles the reference implementation using Nsys to identify performance bottlenecks and generates a structured to-do list of optimization tasks.
\textcircled{2} The Coder executes one subtask per iteration,
\textcircled{3} compiles the modified code with \texttt{nvcc}, and produces an executable for unit testing.
\textcircled{4} The Verifier selects the appropriate profiling tool (Nsys, NCU, or both) to analyze the target kernel.
\textcircled{5} If compilation/correctness fails, the profiling stage is skipped.
\textcircled{6} Relevant  Nvidia's official documentation is retrieved via a RAG module, and
\textcircled{7} Construct a single task feedback to Coder to initiate the next coding-feedback iteration.}
    \label{fig:main}
    \vspace{-5pt}
\end{figure*}
\section{Framework Design and Methodology}
To effectively scale multi-agents system to end-to-end GPU programming, we present StitchCUDA, a multi-agent framework composed of three agents. Planner, Coder, and Verifier that jointly perform iterative optimization of end-to-end GPU programs. The Coder agents are further enhanced with rubric-based agentic RL to improve their responsiveness to feedback from Verifier, enabling more effective self-correction and optimization over iterations. In this section, we will discuss in detail our multi-agent design and the method in rubric-based agentic RL for Coder.

\subsection{Multi-agent Framework Workflow}
\label{sec:multi_agent_workflow}
StitchCUDA orchestrates specialized agents with a global state machine that shares a typed \texttt{State}, containing generated code, profiling artifacts, and routing decisions. The workflow is an Iterative Coding-Feedback Loop, as shown in Fig.\ref{fig:main}. The details of each agent are below:

\vspace{-10pt}
\begin{itemize}
\item \textbf{Planner}: parses the reference PyTorch code, builds a minimal profiling harness, and records Nsys traces to identify time-dominant kernels and system hotspots. The agent then emits a structured to-do list with task identifiers, target kernels, expected shapes, and constraints (e.g., data types or numeric tolerance) that guide downstream code generation. Using chain-of-thought prompting, the Planner reasons at the system level to decompose the workload into subtasks that jointly consider kernel-level efficiency and host-side orchestration, improving the feasibility and coherence of the generated plan.

\item \textbf{Coder}: generates CUDA implementations for the current subtask in a self-contained project (source, build files, Pybind interface) and invokes
\texttt{nvcc} to compile. It writes kernel stubs and launch code, host-side orchestrations that match the Planner's spec, and captures build artifacts (executable path, logs, and generated source) in the shared \texttt{State}. After receiving the Verifier's feedback with optimization suggestions, the Coder refines the current subtask accordingly.

\item \textbf{Verifier}: validates correctness and performance. When compilation fails, it analyzes the error log and returns concrete fix guidance for the Coder (e.g., missing headers, signature mismatches, or type errors). When tests pass, the Verifier analyzes the end-to-end program from two perspectives, using Nsys first identifies the dominant GPU kernel that costs the most GPU time and dominant system-level bottlenecks (e.g., data transfer between CPU and GPU, kernel launch, CPU-GPU synchronization). It then profiles the identified bottleneck kernel with NCU. First, it classifies it as memory-bound or compute-bound and automatically selects relevant performance metrics for detailed analysis. Based on this two-level diagnosis, the Verifier produces actionable optimization suggestions addressing both kernel-level improvements and required system-level changes, which are routed back to the Planner or Coder.
\end{itemize}

In addition, we prepare several up-to-date CUDA/GPU documents (e.g., the white paper of Hooper and Blackwell GPUs, the newest tutorial of CUBLAS/CUTLASS library) for Planner and Verifier to use Retrieval-Augmented Generation \cite{lewis2021retrievalaugmentedgenerationknowledgeintensivenlp} in inference, so that they can provide effective plans and optimization feedback, combining with the newest hardware specs and software stacks. Details about RAG implementations are shown in Appendix~\ref{app:rag}.

Verifier routes the whole loop based on tests' outcomes: attempt the task again if failed, further optimize current task for better performance if successful, then advance to the next task on success and good speedup; replan when fundamental mismatches are detected, overall final test performance is slower than reference; stop when the iteration budget is exhausted. The detailed prompts for every agent are shown in Appendix~\ref{app:prompt}.

\begin{figure*}[ht]
    \centering
    \includegraphics[width=\linewidth]{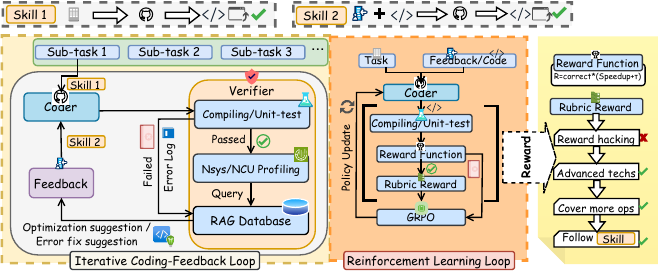}
    \caption{Rubric-based Agentic RL process for Coder. We train the Coder on two atomic skills via GRPO. Besides the rule-based reward function, we introduce a rubric reward to avoid reward hacking and encourage more kernel optimizations.}
    \label{fig:RL}
    \vspace{-10pt}
\end{figure*}

\subsection{Rubric-based Agentic RL for Coder}

\subsubsection{Decomposing Multi-turn Interactive Trajectories into Atomic Skills}

To retain the benefits of agentic RL while avoiding prohibitive multi-turn
rollout cost (See~\ref{app:time}), we decompose the multi-turn agentic RL into learning two
\emph{atomic skills} that recur throughout our workflow (Fig.~\ref{fig:RL}):
\textbf{Skill~1}, generating CUDA kernels from reference PyTorch code and subtask
requirements; and \textbf{Skill~2}, improving an existing kernel by following
feedback and implementing targeted optimizations. We then train these skills in
a single-turn RL.

Concretely, for Skill~1, we sample 80\% of tasks from KernelBench Levels~1--3 and
use a Planner to produce subtasks. After filtering with human
experts, we obtain 200 curated samples containing reference PyTorch code, subtask
requirements, and prompt templates. For Skill~2, we run our framework on the
same tasks, using Qwen3-32B as the Coder to collect verifier feedback, and retain
only feedback instances that lead to a correct next-iteration kernel. This
yields 200 additional samples containing previous code, feedback, and reference PyTorch code. Finally, we merge those samples and apply GRPO to jointly
optimize the Coder across the two skills, providing an efficient surrogate for
multi-turn RL without high rollout overhead.

\subsubsection{Using Rubric Reward for RL Training}

To mitigate reward hacking/degenerate behaviors and encourage more optimization,
we introduce a \emph{Rubric Reward} that provides a comprehensive
assessment of candidate kernels. Rubric-based rewards are commonly used to
evaluate attributes that are not fully captured by simple tests or scalar
metrics~\cite{huang2025reinforcementlearningrubricanchors,liu2026openrubricsscalablesyntheticrubric}.
Our rubric is designed and validated by human CUDA experts with assistance from
advanced LLMs, and scores each candidate along four dimensions: \textbf{(i)
Anti-Hacking}, penalizing reward exploitation; \textbf{(ii) CUDA Engineering},
rewarding advanced optimization techniques; \textbf{(iii) Operator Coverage},
encouraging broader optimization for complex multi-operation programs; and \textbf{(iv)
Skill Compliance}, enforcing adherence to task requirements (Skill~1) or feedback
instructions (Skill~2). Each dimension uses a discrete scale with detailed
criteria per score level. Concretely, we aggregate these scores into a
normalized shaping term:
\begin{equation}
\hat{r}_{\mathrm{rubric}}
=
\left[
\frac{
\sum_{k=1}^{K} s_k- K\,s_{\min}
}{
K\,(s_{\max}-s_{\min})
}
-\frac{1}{2}
\right],
\label{eq:rubric_reward}
\end{equation}
where $s_k \in \{1,\dots,5\}$ denotes the discrete score for the $k$-th rubric
criterion, $K$ is the number of rubric dimensions, and $s_{\min}=1$ and
$s_{\max}=5$ are the per-dimension bounds. This normalization maps the rubric
scores to a centered range, producing a stable shaping signal that complements
the rule-based reward. Details of reward design are in the Appendix~\ref{app:rubric}.

\paragraph{Final reward formulation.}
We combine rubric-based shaping with a rule-based reward function(based on correctness and speedup) to form the final training reward:
{\footnotesize
\begin{equation}
R
=
\mathbb{I}_{\mathrm{corr}}
\cdot
\bigl(1-\mathbb{I}_{\mathrm{hack}}\bigr)
\cdot
\min\!\left(
\left(s + \tau \right)
\left( 1 + \lambda\,\hat{r}_{\mathrm{rubric}} \right),
\; R_{\max}
\right),
\label{eq:final_reward}
\end{equation}
}
where $\mathbb{I}_{\mathrm{corr}} \in \{0,1\}$ indicates whether the generated
kernel passes the benchmark’s functional correctness verification.
$\mathbb{I}_{\mathrm{hack}} \in \{0,1\}$ flags major reward-hacking behaviors;
When $\mathbb{I}_{\mathrm{hack}}=1$, the reward is suppressed to prevent policy
updates from reinforcing exploitation. The scalar $s$ denotes the measured
runtime speedup over the reference implementation. The constant $\tau=0.3$
provides an additive offset to avoid vanishing rewards for correct kernels with
marginal speedup. The coefficient $\lambda=1$ controls the contribution of
rubric-based shaping relative to rule-based reward, and $R_{\max}=5$
caps the maximum reward magnitude to stabilize policy optimization and avoid training collapse.

\paragraph{Training details.}
We train the Coder under a single-turn setting using GRPO, targeting both
skills as described in the previous section. We adopt Qwen3-32B as the Coder backbone, as its strong reasoning capability alleviates sparse-reward and better matches the distribution of RL samples. In each training step, we sample a batch of 16 prompts and generate $8$ rollouts per prompt. We then compute the GRPO objective from~\cite{Guo_2025} and update the policy:

{\tiny
\begin{equation}
\resizebox{\linewidth}{!}{$\displaystyle
    \begin{aligned}
        \mathcal{J}_{\text{GRPO}}(\theta) 
        &= \mathbb{E}_{\substack{q \sim P(Q),  \{o_i\}_{i=1}^G \sim \pi_{\theta_{old}}(O|q)}} 
        \Bigg[ \frac{1}{G} \sum_{i=1}^G \frac{1}{|o_i|} \sum_{t=1}^{|o_i|} \\
        &\quad \min \bigg( 
            \frac{\pi_\theta(o_{i,t} | q, o_{i,<t})}{\pi_{\theta_{old}}(o_{i,t} | q, o_{i,<t})} \hat{A}_{i,t}, \\
        &\quad \text{clip} \left( 
            \frac{\pi_\theta(o_{i,t} | q, o_{i,<t})}{\pi_{\theta_{old}}(o_{i,t} | q, o_{i,<t})}, 1 - \epsilon, 1 + \epsilon 
        \right) \hat{A}_{i,t} 
        \bigg) \\
        &\quad - \beta \mathbb{D}_{KL} \left[ \pi_{\theta} \parallel \pi_{ref} \right] \Bigg],
    \end{aligned}$}
    \label{eq:GRPO-obj}
\end{equation}
}
where $\hat{A}_{i,t} = \frac{r_i - \mathrm{mean}(\mathbf{r})}{\mathrm{std}(\mathbf{r})}$,
$\mathbf{r}=\{r_i\}_{i=1}^{G}$ are the rollout-level rewards, and each $r_i$ is
computed using Eq.~\ref{eq:final_reward}, where we deploy Qwen3-32B to assign rubric reward. Training is conducted on 4 H200 GPUs. We set \texttt{max\_response\_length} to 16384 and use LoRA for compute-efficient fine-tuning with \texttt{rank}$=$\texttt{alpha}$=$128. The sampling parameters follow the Qwen3 Team~\cite{qwen3technicalreport}. Training one Coder costs approximately 20 Hours.

\label{RLdesign}
\section{Experiments and Evaluation}
We evaluate our method on KernelBench\cite{kernelbench}, a popular automated GPU program generation benchmark with 4 Levels. Levels 1 and 2 consist of kernel-level tasks: a single batched matrix multiplication kernel on Level 1 and a single Python file with Conv2D, ReLU, and BiasAdd on Level 2. Level 3 contains 50 end-to-end GPU programming tasks, such as a full VisionTransformer inference implementation with all kernel API calls (e.g., conv2d, relu, maxpool, attention). We identify limitations in KernelBench and address them through modifications; details are provided in Appendix~\ref{app:bench}.

The inputs are generated randomly in various numerical values and shapes for each task. All the reference PyTorch codes in KernelBench are run in eager mode without enabling \texttt{torch.compile}. For Level 3, we also provide an additional comparison with \texttt{torch.compile} enhanced reference code in the main result section. We do not apply it in the reference code for Level 1/2, given its minor impact on performance in single-kernel tasks.

We use Qwen3-32B \cite{qwen3technicalreport,hui2024qwen2} as our RL base model for Coder in our multi-agent framework and GPT-5.2 as the Planner and Verifier. 
 
We evaluate our method and baselines at KernelBench levels 1, 2, and 3 on 2 of the most advanced GPUs: the NVIDIA H200 (Hooper Architecture) and RTX PRO 6000 (Blackwell Architecture), separately. To fully evaluate the effectiveness of StitchCUDA, we apply various baselines as shown in Table~\ref{tab:l3}, including general and domain-specific LLMs, their performance as the Coder in our multi-agent framework, and the only open-source multi-agent framework, CUDAForge~\cite{cudaforge}, using GPT-5.2 as Coder and Judge.
We enable thinking for all LLMs in our experiments and evaluate all baselines and StitchCUDA with 15 iterations in 3 metrics: 
\vspace{-10pt}
\begin{itemize}
    \item \textit{Success Rate}, is measured by checking whether the generated code’s output falls within a predefined error tolerance of the reference implementation across multiple random seeds. A task is deemed successful if the tolerance is met in at least one iteration, independent of speedup. All successful cases are manually reviewed to filter out hacking results (e.g., PyTorch-only code). 
    \item \textit{E2E Average Speedup}, is computed as the ratio between the average end-to-end runtime(including data movement, kernel launch, CPU-GPU synchronization, and GPU kernel time) of the reference code and the generated code for each task. Following \cite{cudaforge}, the reported runtime for the generated code is the best result obtained over 15 refinement iterations.
    \item $\rm{Fast}_1$, A standard metric used in KerneBench ~\cite{kernelbench} that accounts for both success rate and achieved speedup, defined as: 
$\rm{Fast}_1=1/N \sum_{i=1}^{N}(correct_i\wedge{speedup_i > 1})$
   
    where N is the number of tasks.
    
\end{itemize}

\begin{table*}[h]
\caption{The success rate and E2E mean speedup relative to PyTorch Eager of various methods on our testing set. The best results are highlighted in \textbf{bold}, the second best is \underline{underline}. StitchCUDA-Q is our multi-agent framework, with the original Qwen3-32B as the Coder. Similarly, StitchCUDA-K is a backend with Kevin32B, StitchCUDA-G is a backend with GPT-5.2 as Coder, and StitchCUDA-C is a backend with Claude-4.6-Sonnet.}
\label{tab:l3}
{\fontsize{7}{8}\selectfont
\begin{tabular}{c|c|ccc|ccc|ccc}
\hline
                                                                                 &                                    & \multicolumn{3}{c|}{Level 1 (20 Tasks)}                       & \multicolumn{3}{c|}{Level 2 (20 Tasks)}                       & \multicolumn{3}{c}{Level 3 (10 Tasks)}                                                           \\ \cline{3-11} 
\multirow{-2}{*}{Hardware}                                                       & \multirow{-2}{*}{Method}           & Correctness    & Avg Speedup    & $\rm{Fast}_1$           & Correctness    & Avg Speedup    & $\rm{Fast}_1$ & Correctness                 & Avg Speedup                  & $\rm{Fast}_1$                   \\ \hline
                                                                                 & GPT5.2                             & 10/20          & 0.73x          & 25\%          & 8/20           & 0.66x          & 30\%             & 2/10                        & 0.47x                        & 10\%                     \\
                                                                                 & Claude-4-sonnet                    & 12/20          & 1.02x          & 25\%             & 8/20           & 0.76x          & 30\%             & 3/10                        & 0.67x                        & 20\%                     \\
                                                                                 & Qwen3-32B                          & 2/20           & 0.02x          & 5\%           & 0            & 0              & 0\%                & 0                           & 0                            & 0\%                     \\
                                                                                 & Kevin32B                           & 9/20           & 0.65x          & 10\%             & 11/20          & 0.49x          & 30\%             & 4/10                        & 0.22x                        & 0\%                        \\
                                                                                 & CUDAForge                          & {\underline{18/20}}    & 2.74x          & {\underline {50\%}}       & \underline{18/20}          & 1.28x          & 80\%             & 6/10                        & 0.75x                        & 30\%                     \\
                                                                                 & StitchCUDA-Q  
                                                                                 & 3/20 & 0.19x & 5\% & \underline{18/20}              & 0.77x              & 45\%                & 6/10	&0.40x	& 10\%                               \\
                                                                                 & \multicolumn{1}{l|}{StitchCUDA-K} & 13/20          & 2.11x          & 30\%          & 17/20          & 0.82x          & 65\%          & 7/10 & { 0.42x} & { 0\%} \\
                                                                                 & \multicolumn{1}{l|}{StitchCUDA-G} & \textbf{19/20} & \underline{3.97x} & \textbf{55\%} & \textbf{19/20} & \underline{1.87x} & \textbf{95\%} & 6/10                  & { 0.99x}                  & {\underline{ 50\%}}            \\
                                                                                 & StitchCUDA-C                            & \underline{18/20}          & \textbf{4.15x}          & \textbf{55\% }         & \textbf{19/20}           & \textbf{1.89x}          & 80\%             & \underline{9/10}                        & \underline{1.18x}                        & \underline{50\%}                     \\
                                                                                 
\multirow{-9}{*}{\begin{tabular}[c]{@{}c@{}}PRO 6000\\ (BlackWell)\end{tabular}} & StitchCUDA                        & {\underline{ 18/20}}    & {\ 2.86x}    & 35\%          & \textbf{19/20} & { 1.55x}   & {\underline{ 85\%}}    & \textbf{10/10}              & \textbf{1.27x}               & \textbf{70\%}            \\ \hline
                                                                                 & GPT5.2                             & 10/20          & 0.6x           & 20\%             & 9/20           & 0.63x          & 30\%             & 2/10                        & 0.48x                        & 10\%                     \\
                                                                                 & Claude-4-sonnet                    & 12/20          & 1.04x          & 30\%             & 10/20          & 0.78x          & 30\%             & 3/10                        & 0.64x                        & 20\%                     \\
                                                                                 & Qwen3-32B                          & 2/20           & 0.21x          & 5\%           & 0              & 0              & 0\%                & 0                           & 0                            & 0\%                        \\
                                                                                 & Kevin32B                           & 9/20           & 0.62x          & 10\%             & 10/20          & 0.53x          & 30\%             & 2/10                        & 0.34x                        & 10\%                        \\
                                                                                 & CUDAForge                          & \textbf{18/20} & 3.18x          & 45\%          & \underline{18/20}          & 1.43x          & 75\%          & 6/10                        & 0.87x                        & 40\%                  \\
                                                                                 & StitchCUDA-Q                      & \underline{17/20}          & 2.13x          & 20\%          & 16/20         & 0.94x          & 50\%          & 3/10                        & 0.24x                        & 10\%                  \\
                                                                                 & \multicolumn{1}{l|}{StitchCUDA-K} & 12/20          & 2.25x          & \underline{50\%}          & 15/20          & 0.77x          & 60\%          & 5/10                        & 0.32x                        & 0\%                        \\
                                                                                 & \multicolumn{1}{l|}{StitchCUDA-G} & \textbf{18/20} & \textbf{3.57x} & 45\%          & \textbf{19/20} & {\underline {1.79x}}    & \textbf{90\%} & {\underline{ 6/10}}                  & {\underline{ 1.01x}}                  & {\underline{ 50\%}}            \\
\multirow{-9}{*}{\begin{tabular}[c]{@{}c@{}}H200\\ (Hooper)\end{tabular}}        & StitchCUDA                        & {\textbf{ 18/20}}    & {\underline{ 3.54x}}    & \textbf{55\%} & \underline{18/20}          & \textbf{1.82x} & {\underline{ 85\%}}    & \textbf{9/10}               & \textbf{1.50x}               & \textbf{70\%}         \\ \hline
\end{tabular}}
\end{table*}

\vspace{-15pt}
\subsection{Main results}
During RL for Coder, we randomly selected 80\% of tasks from Levels 1/2/3 to collect training data. We evaluate our method and all baselines on the remaining 20\% of tasks as the test set. The experiment results are shown in Table~\ref{tab:l3}. Across both GPUs, the results support the following claims: 
\paragraph{Multi-agent orchestration substantially improves end-to-end correctness and makes speedups attainable.}

We first evaluate the contribution of multi-agent orchestration by holding the Coder model fixed and only changing the workflow from single-shot generation to StitchCUDA’s multi-agent framework.
Using Qwen3-32B as an example (our RL base model), single-shot generation performs poorly in the end-to-end setting: on H200, Qwen3-32B achieves only \textbf{2/20} correctness on Level~1 and no successful cases on Level~2/3. In contrast, StitchCUDA-Q (same Qwen3-32B as Coder, \emph{no RL}) increases Level~1 correctness to \textbf{17/20} with \textbf{2.13$\times$} mean speedup, and makes harder levels attainable (Level~2: \textbf{16/20} correctness; Level~3: \textbf{3/10} correctness).
A similar trend holds for a stronger domain-specific RL model: Kevin32B alone remains limited on Level~3 (RTX PRO 6000: \textbf{4/10}, \textbf{0.22$\times$}; H200: \textbf{2/10}, \textbf{0.34$\times$}), while placing Kevin32B into our multi-agent workflow (StitchCUDA-K) substantially improves correctness on harder benchmarks (RTX PRO 6000 Level~3: \textbf{4/10}$\rightarrow$\textbf{7/10};  H200 Level~3: 2/10 $\rightarrow$ 5/10). Even for advanced GPT-5.2 and Claude-4.6-Sonnet, the trend holds the same(GPT-5.2 v.s. StitchCUDA-G and Claude-4.6-Sonnet v.s. StitchCUDA-C).
These comparisons indicate that our multi-agent decomposition and tool-augmented iteration are essential, improving the performance of various models.

\paragraph{Agentic RL strengthens the multi-agent loop into consistent optimization on end-to-end tasks.}
We evaluate the impact of agentic RL by comparing StitchCUDA against its no-RL variant, StitchCUDA-Q. RL yields a dominant gain on the hardest Level~3 tasks. On H200 GPU, correctness improves from \textbf{3/10} to \textbf{9/10}, mean speedup increases from \textbf{0.24$\times$} to \textbf{1.50$\times$}, and $\rm{Fast}_1$ rises from \textbf{10\%} to \textbf{70\%}. The same pattern holds on Level~2 (from \textbf{16/20}, \textbf{0.94$\times$}, \textbf{50\%} $\rightarrow$ \textbf{18/20}, \textbf{1.82$\times$}, \textbf{85\%}) and Level~1 (same corrcetness with better speedup and $\rm{Fast}_1$). And this trend holds the same on RTX PRO 6000. This shows that our rubric-based agentic RL is the key factor driving significant real-system-level speedups, rather than merely improving correctness.
\paragraph{Agentic RL gains persist beyond using a stronger model.}
Even compared to StitchCUDA-G (GPT-5.2 as all agents), StitchCUDA achieves substantially stronger Level~3 results on both GPUs, with a much smaller 32B Coder model. StitchCUDA and StitchCUDA-G gain similar performance on Level~1/2, with nearly 100\% correctness and best or second-best mean speedup. On Level~3, StitchCUDA surpasses StitchCUDA-G on all metrics, and is the only method that can achieve nearly 100\% correctness on Level~3.

We evaluate StitchCUDA, StitchCUDA-G, and Kevin-32B on the Level 3 test set with reference code enabling \texttt{torch.compile} on H200 GPU. Kevin32B achieves lower speedup from 0.34$\times$ in eager mode compared to 0.18$\times$ in \texttt{torch.compile} enabled reference code. StitchCUDA-G achieves 0.92$\times$ speedup, lower than its 1.01$\times$ in eager mode. StitchCUDA achieves 1.29$\times$, also lower than 1.50$\times$ in eager mode, but still better than the reference code. 
We want to emphasize that \textbf{our agents manually perform system optimization as a result of improvement over \texttt{torch.compile}, such as custom kernel fusion and data movement optimization, without introducing compiler overhead}.

\label{sec:main_result}
\subsection{Ablation Study}
\begin{figure}[t]
    \centering
    \includegraphics[width=\linewidth]{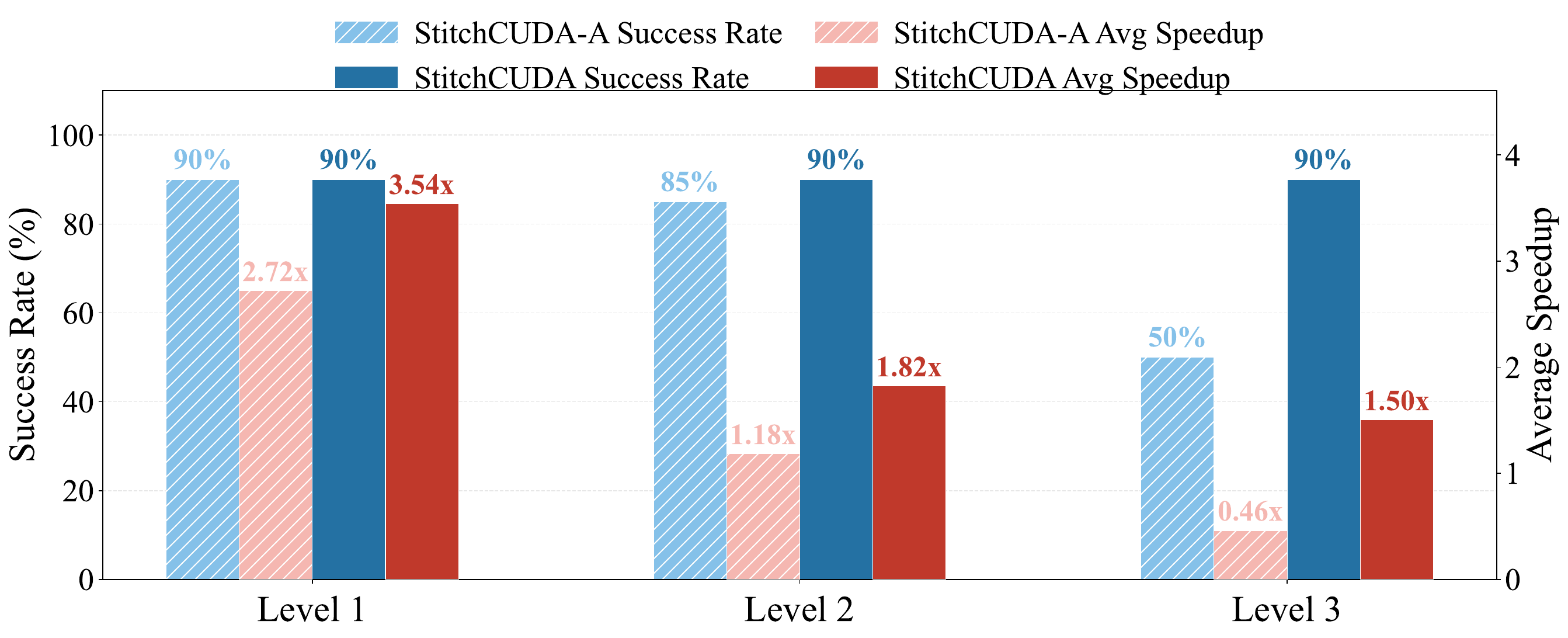}
    \caption{Comparison of success rate and average speedup between StitchCUDA and StitchCUDA-A across Level 1/2/3. StitchCUDA
  achieves higher success rates and speedups at all levels, with the rubric-based reward.}
    \label{fig:ablation}
    \vspace{-10pt}
\end{figure}

To comprehensively evaluate the effect of the rubric reward, we conduct an ablation study. 
We introduce StitchCUDA-A, an ablated variant of StitchCUDA
trained with \emph{only} the rule-based reward (correctness and speedup; see
Appendix~\ref{kevinreward}) using Qwen3-32B. As shown in Fig~\ref{fig:ablation}, we evaluate StitchCUDA-A
on KernelBench Levels~1/2/3. Relative to StitchCUDA,
StitchCUDA-A achieves a comparable success rate on Levels~1/2, but a
lower success rate on Level~3 and lower speedup across Levels~1/2/3, indicating
that rubric reward is critical for effective RL training. We further inspect
Level~3 failures to understand the role of rubric reward and observe three
recurring patterns: \textbf{(1) Degenerate solutions.} The model often
produces overly conservative, low-performance implementations; \textbf{(2) Weak
feedback following.} Without explicit reward, the model’s ability to act on feedback degrades; and \textbf{(3) Reward hacking.} In the absence of reward penalties, the model exhibits reward hacking similar to Kevin-32B. Additional analysis for reward-hacking is provided in Section~\ref{sec:hack}.

We also conducted a per-dimension ablation study of the 4 aspects of our rubrics. We train four variants, each removing one rubric dimension while keeping the other three, and evaluated on KernelBench Level~3 on H200 GPU. 

As shown in Table~\ref{tab:rubric}. The results reveal that each dimension contributes distinctly. Removing Anti-Hacking has the most severe overall impact: correctness drops sharply (9.0 → 5.7), hacking instances nearly double (7.7 → 15.7), and speedup degrades to 0.75×. This indicates that without explicit anti-hacking reward, the model extensively exploits reward loopholes, producing solutions that are neither correct nor genuinely optimized. This dimension serves as the foundation of the entire rubric system. Removing CUDA Engineering maintains reasonable correctness (8.0) and low hacking (9.3), but reduces speedup from 1.50× to 1.18×, confirming that this dimension drives genuine optimization quality by encouraging advanced techniques such as tiling, tensor cores, and kernel fusion. Removing Operator Coverage shows a similar pattern with even lower speedup (1.08×), as the model reverts to optimizing only trivial operators while leaving core bottlenecks untouched. Removing Skill Compliance primarily affects correctness (9.0 → 7.3) while maintaining moderate speedup (1.33×), indicating that the model retains optimization capability but fails to follow task specifications and Verifier feedback precisely. All single-dimension ablations outperform StitchCUDA-A (no rubric), confirming that the remaining three dimensions still provide meaningful training signal.

\begin{table}
    \centering
    \caption{Ablation results of training rubrics}
{\fontsize{8}{8}\selectfont
    \begin{tabular}{|c|c|c|c|}
\hline
\textbf{Variant}      & \textbf{Corre. (/10)} & \textbf{Speedup} & \textbf{Hacking (/50)} \\ \hline
StitchCUDA            & 9.0                        & 1.50×            & 7.7                             \\
w/o Anti-Hacking      & 5.7                        & 0.75×            & 15.7                            \\
w/o CUDA Engineering  & 8.0                        & 1.18×            & 9.3                             \\
w/o Operator Coverage & 8.0                        & 1.08×            & 10.0                            \\
w/o Skill Compliance  & 7.3                        & 1.33×            & 8.7                             \\
StitchCUDA-A          & 5.0                        & 0.46×            & 16.3                            \\ \hline
\end{tabular}}
    \label{tab:rubric}
\end{table}

\subsection{Hacking detection}
\label{sec:hack}
In our evaluation, we observe serious hacking behaviour on RL-based Kevin-32B, including writing PyTorch-only code or hardcoding output. We try to address it by using prompt engineering and test if the problem can be solved. Specifically, we explicitly add the following to the prompts: “You must at least customize one kernel in your implementation, and do not copy the original PyTorch API. You cannot hardcode output.” However, the problem remains. To gain a comprehensive understanding, we examine hacking behavior across the evaluations of the following 4 methods: StitchCUDA-K, StitchCUDA-Q, StitchCUDA-A, and StitchCUDA. Therefore, we can examine how RL influences hacking behaviors and the impact of the rubric reward, including the hacking penalty.

Hacking detection is performed on our test sets across Levels~1/2/3, and the results are shown in Table~\ref{tab:hack}. We count two kinds of hacking:  \textbf{partial hacking} and \textbf{total hacking}. The former means the model exhibits hacking behavior in the 15 rounds' iteration, while the latter means all 15 results exhibit hacking behavior. The hacking behavior is counted by CUDA experts, containing “PyTorch-only code” and “hardcode output”.

According to our results, even with explicit prompts, StitchCUDA-K using RL-based Kevin-32B exhibits 22 partial hackings in the whole test and gets 4 total hackings, significantly worse than StitchCUDA-Q with  Qwen3-32B. However, another RL-based StitchCUDA-A performs 10 partial hackings but do not exhibit any total hacking. We attribute it to  our RL trains the model to follow feedback, where we ask the model not to hack. In contrast, StitchCUDA performs the lowest hacking rate (8 partial hackings) in the test, demonstrating the effectiveness of rubric reward for anti-hacking.

\begin{table}[h]
    \centering
    \tiny
    \caption{Hacking detection with prompts hint.}
    \label{tab:hack}
    {\fontsize{6}{7}\selectfont
    \begin{tabular}{l|cccccccc}
        \hline
        & \multicolumn{2}{c}{Level 1} & \multicolumn{2}{c}{Level 2} & \multicolumn{2}{c}{Level 3} & \multicolumn{2}{c}{Sum} \\
        \cline{2-9}
         Model& partial & total & partial & total & partial & total & partial & total \\
        \hline
        StitchCUDA-K     & 10 & 3 & 8 & 0 & 4 & 1 & 22/50 & 4/50 \\
        StitchCUDA-Q     & 4  & 0 & 4 & 0 & 2 & 0 & 10/50 & 0   \\
        StitchCUDA-A & 8  & 0 & 5 & 0 & 3 & 0 & 16/50 & 0   \\
        StitchCUDA & 3  & 0 & 3 & 0 & 2 & 0 & 8/50  & 0   \\
        \hline
    \end{tabular}}
\end{table}

\section{Discussion}
\label{sec:discuss}
In this section, we firstly discuss two observations in preventing reward hacking and LLM's degenerate behaviors in complex end-to-end GPU programming. Then we also discuss current limitation and possible future directions.

\subsection{Format check fails to prevent reward hacking}
In our early experiment, we try to use format check to prevent reward hacking in RL, which is also applied by Kevin-32B~\cite{kevin}. However, format check fails to prevent reward hacking, both in our experiments and Kevin-32B's result in Table~\ref{tab:hack}. We attribute the failure to two reasons: First, some reward hacking behaviors are very subtle and hard to be detected by a regex-based scripts, such as hard coding. Second, format check faces a trade-off, where strict checks could classify correct solution as “hacking”, and loose checks could omit some hacking solutions. Therfore, format check alone is not a good remedy for avoiding reward hacking on KernelBench.

\subsection{Degenerate behaviors on KernelBench}
We also observe that current LLMs clearly exhibit degenerate behaviors in complex end-to-end GPU programming tasks. For example, in many Level 2/3 tasks with multiple kernels, CUDAForge and Kevin-32B show a strong tendency to only modify “easy parts,” with only minor impact on performance. For example, only write a ReLU kernel and leave all others (e.g., conv2d and GEMM) unchanged.  And they will even replace only one of multiple ReLU kernels in the reference code with a customized kernel. In this case, the program can be correct with a very minor different performance (typically less than 5\% in our experiments). A case of this kind of code generated by Kevin is shown in Appendix~\ref{app:fear}. We attribute the degenerate behaviors to two reasons:

\textbf{RL reward design pushes the model.} According to the rule-based reward design of Kevin-32B in Appendix~\ref{kevinreward}. A correct but low-speed kernel will get higher rewards than a kernel with aggressive optimizition but has a minor error, hindering model try those advanced optimization techniques. See Appendix~\ref{app:rubric} for how StitchCUDA address it.

\textbf{Current LLMs have no confidence.} We manually track the chain of thought of these models in thinking mode to figure out the reason of degenerate behaviors. We observe a lot of similar content in their chain of thought, such as “current conv2d in PyTorch is backend with cuBLAS lib, which is hard to surpass, focus on ReLU and Linear kernels first.”. This reveals that current LLMs have no confidence to challenge those real critical parts in GPU programming. And as a result, they also lose the opportunity to utilize low-level deep optimization techniques in practice, such as fine-grained tiling and tensor cores. Based on our observations, Kevin32B never use these advanced techniques in experiments, which partially explains why its speedup decreases significantly on KernelBench level 3. 

\subsection{Limitation and future directions}
Although \textsc{StitchCUDA} demonstrates strong performance on end-to-end GPU programming tasks, several important limitations remain for this area.

\textbf{Limited generalization across hardware generations.}
Our experiments cover H200 and RTX PRO 6000, but GPU optimization is highly architecture-dependent. Code that performs well on one platform may not transfer to another due to differences in hardware specs such as shared memory capacity and tensor core support.  Therefore, it remains unclear whether an LLM-generated GPU program can consistently preserve correctness and performance across hardware generations.

\textbf{Adapting LLMs to rapidly evolving GPU knowledge.}
\textsc{StitchCUDA} uses RAG to provide agents with up-to-date CUDA/GPU documentation, which partially mitigates the knowledge staleness of pretrained LLMs. However, large-scale LLM pretraining  are difficult to keep synchronized with the fast iteration of GPU architectures stacks. New hardware features, programming interfaces, and optimization practices may appear after a model is trained, and simply retrieving documentation does not guarantee that the model can correctly interpret such knowledge and apply it to CUDA programming decisions. How to make LLM agents effectively retrieve and reason over the latest hardware knowledge remains an important future direction.

\textbf{Limitations of current benchmarks.}
Current CUDA-generation benchmarks~\cite{kernelbench,zhu2026cudabench,li2025tritonbench} mainly evaluate correctness and speedup under a small set of fixed environments. They do not systematically measure cross-platform portability, robustness to software-stack changes, deployment safety, or long-term regression behavior. Deployment-quality GPU software should also be evaluated under real application semantics, such as training stability for ML training workloads, rather than only output correctness on isolated benchmark inputs. Future benchmarks should therefore move beyond single-environment performance evaluation and incorporate portability, robustness, safety, and regression testing. At the time of this paper's publication, several concurrent and recent benchmark efforts have started to move in this direction, highlighting the growing need for more robust evaluation of LLM-generated GPU programs~\cite{computeeval,cudahercules,cudabeaver}.

\section{Conclusion}
In this paper, we propose StitchCUDA, a multi-agent framework that integrates Rubric-based reinforcement learning with the Coder agent. We identify three main challenges in LLM-based automated end-to-end GPU program generation. We improve the multi-agent framework through well-designed multi-agent decomposition and fundamentally enhance the Coder's CUDA coding ability through rubric-based agentic reinforcement learning. Comprehensive experiment results show that StitchCUDA delivers excellent performance on end-to-end GPU programming tasks, significantly outperforming simple multi-agent approaches and RL models.
\section*{Acknowledgements}
This research was supported in part by NSF CIF-2414372 and Amazon.


\section*{Impact Statement}


This paper presents work whose goal is to advance the field of Machine
Learning. There are many potential societal consequences of our work, none
which we feel must be specifically highlighted here.



\bibliography{ref}
\bibliographystyle{icml2026}

\newpage
\appendix
\onecolumn
\section{KernelBench Modification.}
\label{app:bench}
We use KernelBench~\cite{kernelbench} as our training environment (80\% tasks) and testing environment(20\% tasks). During training, we identified limitations in the original KernelBench and addressed them manually. These issues can lead to misleading results in evaluation, we show an example as following.

Level2 task 80, which contains 4 kernels. The original model applies a linear projection (GEMM via \texttt{nn.Linear}) to the input, then reduces by taking the maximum along a specified feature dimension (with keepdim=True, yielding a single value per sample when reducing over the feature axis). It then mean-centers the reduced tensor by subtracting the mean computed along dimension 1, and finally applies a GELU to produce the output.

However, after the max-reduction, the tensor collapses from shape $(B, O)$ to $(B,1)$:
\[
x = \max_{j \in \{1,\dots,O\}} \tilde{x}_{:,j}, \qquad x \in \mathbb{R}^{B \times 1},
\]
where $\tilde{x} \in \mathbb{R}^{B \times O}$ is the output of the linear layer.

For any sample $i \in \{1,\dots,B\}$, the $i$-th row contains a single scalar,
\[
x_i = [m_i] \in \mathbb{R}^{1}.
\]
Therefore, the mean over dimension $1$ equals the element itself:
\[
\mathrm{mean}(x_i) = m_i.
\]
The subsequent mean-centering step becomes
\[
x_i - \mathrm{mean}(x_i) = m_i - m_i = 0,
\]
so the tensor is identically zero:
\[
x - \mathrm{mean}(x, \mathrm{dim}=1) = \mathbf{0} \in \mathbb{R}^{B \times 1}.
\]
Finally, since $\mathrm{GELU}(0)=0$, the output remains all zeros, no matter what inputs and what size of inputs. 

In this case, even the Coder gives a fault code and does not implement any of 4 kernels in this model, the output can be just correct because the correct answer is always a tensor with all 0 values. We even find that sometimes Planner can identified this bug and just tell Coder to implement a “fill-zero kernel” that just fills a tensor with 0 value. And the Verifier will eventually report a significant speedup. 

In our experiments, we replace the hard $\max$ reduction with a $\mathrm{top}\text{-}k$ selection, so the model retains the $k$ largest feature values per sample instead of collapsing the feature dimension to a single scalar. Concretely, it changes the computation graph from
\[
\text{GEMM} \;\rightarrow\; \max(\mathrm{dim}=1) \;\rightarrow\; \text{mean-center on } \mathrm{dim}=1 \;\rightarrow\; \mathrm{GELU}
\]
to
\[
\text{GEMM} \;\rightarrow\; \mathrm{top}\text{-}k(\mathrm{dim}=1) \;\rightarrow\; \text{mean-center on } \mathrm{dim}=1 \;\rightarrow\; \mathrm{GELU}.
\]

This avoids the original issue because after $\mathrm{top}\text{-}k$, each row has $k>1$ elements (i.e., $x \in \mathbb{R}^{B \times k}$). Hence, the per-row mean is no longer trivially equal to the single retained element, and the mean-centering step does not degenerate to $x - x = 0$. Formally, for sample $i$,
\[
x_i = [v_{i,1}, \ldots, v_{i,k}] \in \mathbb{R}^k,\quad
x_i - \frac{1}{k}\sum_{\ell=1}^k v_{i,\ell} \neq \mathbf{0}
\]
In general, the output will not deterministically be all zeros.
Similar issues have been identified in the following tasks: Task 12, 85, 87 on Level 1; Task 80, 83 on Level 2; and Task 33, 41, 45, 50 on Level 3. We fix all of them manually before we collect RL training data and our evaluation.

\section{Additional details in multi-agent framework}
\subsection{Prompts and example outputs}
\label{app:prompt}
Here we show the prompts for our three agents separately. 

\begin{tcolorbox}[mybox, title={Prompts for Planner}]
\ttfamily
\tiny
\begin{verbatim}
You are an expert CUDA performance engineer. 
Your task is to analyze the PyTorch reference code and create an optimized CUDA implementation plan.

## Goal
Create a high-performance CUDA/C++ program that:
1. Implements the SAME FUNCTIONALITY as the PyTorch reference (correctness is mandatory)
2. Maximizes GPU utilization and minimizes latency
3. Uses advanced optimization techniques (tiling, fusion, tensor cores, etc.)

## Target Hardware
**GPU:** $gpu_name (sm_120 architecture - Blackwell)
- Tensor cores available (use XMMA/WMMA for matrix operations)
- High memory bandwidth - optimize for coalesced access
- Large shared memory - use for data reuse

## PyTorch Reference Code
```python
$reference_code
```

## NSYS Profiling of Reference
$profiling_summary

## Your Task

Analyze the reference code and create an optimization plan:

1. **Understand the computation graph** - what operations are performed and in what order
2. **Identify optimization opportunities**:
   - Which operations can be fused? (e.g., conv+bias+relu, matmul+add)
   - Which operations benefit from tensor cores? (matmul, conv with specific shapes)
   - Where is memory bandwidth the bottleneck?
   - What tiling strategies work best for the data sizes?
3. **Plan the implementation** - specify each kernel with optimization details

## Output JSON Format

```json
{
  "project_name": "cuda_project",
  "analysis": {
    "operations": ["list all operations from reference code"],
    "input_shapes": ["exact shapes from get_inputs()"],
    "output_shape": "exact output shape",
    "batch_size": "from reference code",
    "total_weights": "total number of weight parameters",
    "weight_specs": [
      {"name": "layer.weight", "shape": [...], "size": N}
    ],
    "fusion_opportunities": [
      {"ops": ["op1", "op2"], "reason": "why fuse"}
    ],
    "intermediate_buffers": [
      {"name": "buffer_name", "shape": [...], "size_bytes": N}
    ]
  },
  "kernel_specs": [
    {
      "function_name": "kernel_name",
      "description": "what this kernel does",
      "inputs": "input tensor specs with exact shapes",
      "outputs": "output tensor specs with exact shapes",
      "parameters": {"all parameters needed for this kernel"},
      "use_library": "cuBLAS | CUTLASS | none",
      "optimization_notes": "specific optimization strategy for this kernel"
    }
  ],
  "execution_order": [
    "1. kernel1: input -> intermediate1",
    "2. kernel2: intermediate1 -> intermediate2",
    "..."
  ],
  "optimization_strategy": {
    "overall_approach": "high-level optimization strategy",
    "memory_optimization": "how to minimize memory traffic",
    "compute_optimization": "how to maximize compute utilization",
    "fusion_plan": "which operations to fuse and why"
  }
}
```

## Optimization Principles

1. **Analyze the specific model** - understand its unique characteristics
2. **Identify bottlenecks** - is it memory-bound or compute-bound?
3. **Plan fusions** - eliminate unnecessary memory round-trips
4. **Choose right tools** - cuBLAS for GEMM, hand-written for custom ops
5. **Use tensor cores** - when matrix dimensions are suitable (multiples of 8)
6. **Optimize memory access** - coalescing, shared memory, tiling

Output only the JSON.
"""
\end{verbatim}
\normalfont
\end{tcolorbox}

\begin{tcolorbox}[mybox, title={Prompts for Verifier}]
\ttfamily
\tiny
\begin{verbatim}
You are a CUDA specialist. Your job is to analyze profiling results and provide optimization feedback.

## Current Code Status
- Iteration: $iteration
- Compiled: $is_compiled
- Correct Output: $is_correct
- Speedup vs Baseline: ${speedup}x

## Generated Code
```python
$generated_code
```

## Your Task

**IMPORTANT: Profiling results (NSYS and NCU) are provided ABOVE this prompt. Analyze them directly.**

### 1. Analyze Profiling Results

### 2. Provide ONE Specific Optimization

Based on the profiling data, identify the PRIMARY bottleneck and suggest ONE concrete fix.

### 3. Provide Verification Feedback
After analysis, provide structured feedback in JSON format:

```json
{
  "verification_status": "pass" | "fail" | "needs_optimization",
  "bottleneck_type": "memory-bound" | "compute-bound" | "low-occupancy" | "stall-issues" | "none",
  "performance_issues": 
    {
      "bottleneck": "Description of bottleneck from profiling",
      "evidence": "Specific metric values (e.g., dram__throughput: 85%)",
      "optimization": "ONE specific optimization to address this"
    },
  "profiling_summary": {
    "nsys_results": "Key findings from NSYS",
    "ncu_results": "Key findings from NCU metrics",
    "primary_bottleneck": "The main performance limiter"
  },
  "files_to_modify": 
    {
      "file_path": "kernel.cu",
      "file_type": "cuda",
      "changes_needed": "Description of what to change"
    },
  "next_steps": 
    "ONE specific actionable optimization",
  "routing_decision": "coding" | "next_task" | "final_test",
  "routing_reasoning": "Brief explanation"
}
```

**Routing Decision Guidelines:**
- **"coding"**: Test failed, compilation error, or correctness issue → retry SAME task (provide feedback)
- **"next_task"**: Test passed and kernel is correct → mark task complete, advance to NEXT task
- **"final_test"**: ALL tasks in todo list completed → run final integration test

**How to decide:**
1. If test bench fails or code has errors → "coding"
2. If test bench passes for current task AND more tasks remain → "next_task"
3. If test bench passes for current task AND this is the LAST task → "final_test"

**Important Guidelines:**
- Be specific: Reference actual code, metrics, and line numbers
- Prioritize: List issues from most to least critical
- Be actionable: Every issue should have a clear suggested fix
- Use profiling wisely: Don't profile if the issue is obvious from code review
"""
\end{verbatim}
\normalfont
\end{tcolorbox}

\begin{tcolorbox}[mybox, title={Prompts for Coder}]
\ttfamily
\tiny
\begin{verbatim}
'''\
You are a senior CUDA kernel optimization specialist.

## MODE: Separate CUDA Files

You are working on a project with SEPARATE CUDA source files, NOT inline code.

## Current Task

**Task Description:** $task_description
**CUDA File:** $cuda_filename
**Kernel Name:** $kernel_name
**Expected Signature:** $kernel_signature
**Data Type:** $dtype
**Implementation Hint:** $implementation_hint

**GPU Specifications:**
$gpu_specs

## Current File Contents

The file `$cuda_filename` currently contains:

```cuda
$current_file_contents
```

## Your Job

Implement ONLY the kernel in `$cuda_filename`. Output:

1. The COMPLETE contents of `$cuda_filename` (with your implementation)
2. NO other files (bindings.cpp, setup.py, etc. are already generated)
3. NO Python code (ModelNew will be auto-generated)

## Implementation Requirements

### CRITICAL: You MUST implement a `__global__` kernel

**ALWAYS write a `__global__` kernel function.** This is required for profiling.

**Prefer custom kernels over library calls.** Custom kernels allow fine-grained optimization and profiling.


**DO NOT** write a file with only host-side library calls and no `__global__` kernel.
**DO NOT write fallback code** that uses generic library functions without optimization.

## Output Format (STRICT)

Output ONLY the CUDA code within code blocks:

```cuda
// Your complete implementation of $cuda_filename
```

NO explanations, NO additional text, ONLY the CUDA code.

Begin!```
\end{verbatim}
\normalfont
\end{tcolorbox}

\subsection{RAG database building}
\label{app:rag}
\begin{figure}[h]
    \centering
    \includegraphics[width=0.8\linewidth]{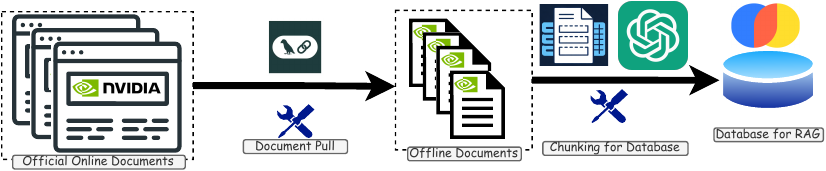}
    \caption{Illustration of online documents processed into the database for RAG}
    \label{fig:rag-database}
    \vspace{-10pt}
\end{figure}

In Fig.~\ref{fig:rag-database}, we curate a set of official NVIDIA CUDA library webpages, including tutorials and documentation for cuBLAS and CUTLASS, to serve as authoritative references for the Planner and Verifier agents. These sources provide up-to-date API specifications and usage guidance to support the Coder. Each webpage is retrieved and converted into text documents using LangChain’s \texttt{WebBaseLoader}. The resulting documents are segmented into overlapping chunks of 1000 tokens with a 100-token overlap to preserve contextual continuity across segments. We then encode each chunk using the \texttt{text-embedding-3-small} embedding model and store the resulting embeddings in a Chroma vector database for retrieval. Table~\ref{tab:rag} presents examples of official documentation and code snippets included in the RAG database.

It is hard to quantitatively evaluate RAG's effects in our framework because we allow our agents to freely choose whether to use it in each inference. Only when the Verifier or Planner thinks it is necessary to search for something in the RAG database will they retrieve it. But we have observed some evidence that implies it is helpful. For example, when the Coder triggers a compile error with a cuBLAS API name, Verifier sometimes searches the RAG database to provide the correct API name, which helps the framework spend less time on minor errors. Without RAG, LLMs can rely only on knowledge gained from pre-training, which may lead to hallucinations during inference.

\begin{table}[h]
\centering
\begin{tabularx}{\linewidth}{|l|X|}
\hline
\textbf{Topic} & \textbf{URL} \\
\hline
CUDA-12.9 & \small \url{https://docs.nvidia.com/cuda/archive/12.9.0/} \\
cuBLAS-12.9 & \small \url{https://docs.nvidia.com/cuda/archive/12.9.0/cublas/index.html} \\
cuSPARSE-12.9 & \small \url{https://docs.nvidia.com/cuda/archive/12.9.0/cusparse/index.html} \\
CUTLASS-4.3.5 & \small \small\url{https://docs.nvidia.com/cutlass/latest/} \\
cuSPARSELt-0.8.1 & \small \small \small\url{https://docs.nvidia.com/cuda/cusparselt/index.html} \\
Nvidia Blackwell Architecture Whitepaper & \small\url{https://images.nvidia.com/aem-dam/Solutions/geforce/blackwell/nvidia-rtx-blackwell-gpu-architecture.pdf} \\
Nvidia Hopper Architecture Whitepaper & \small\url{https://resources.nvidia.com/en-us-hopper-architecture/nvidia-h100-tensor-c} \\
Blackwell SM100 GEMMs APIs & \small\url{https://docs.nvidia.com/cutlass/latest/media/docs/cpp/blackwell_functionality.html} \\
Nsys Documentation & \small\url{https://docs.nvidia.com/nsight-systems/UserGuide/index.html} \\
NCU Profiling Guide & \small\url{https://docs.nvidia.com/nsight-compute/ProfilingGuide/index.html}\\
\hline
\end{tabularx}
\caption{Example official documentation topics and links}
\label{tab:rag}
\end{table}

\section{Additional details in rubric-based agentic reinforcement learning}
\label{app:rl}
\subsection{Challenges of RL for CUDA generation and rubric reward design}

Existing RL-based methods are vulnerable to reward hacking and degenerate behaviors. In this section, we will provide some cases of reward hacking/degenerate behaviors and demonstrate how we use the rubric reward to mitigate these challenges.

\subsubsection{Challenges of reward hacking}
\label{app:rewardhack}
Reward hacking refers to behaviors that maximize a specified reward by exploiting loopholes in the reward or evaluation procedure, yielding high reward but poor alignment with the intended task objective. In the domain of CUDA generation, a common rule-based reward design is the following formulation from Kevin-32B~\cite{kevin}, combining functional correctness and speedup:
$$
\label{kevinreward}
S = 0.3 \cdot \mathbf{1}_{\{\text{correct}\}} + \frac{T_{\text{baseline}}}{T_{\text{kernel}}}  \cdot \mathbf{1}_{\{\text{correct}\}}
$$
Functional correctness is checked against the reference PyTorch implementation when tested with randomized inputs; speedup is computed as the ratio of the PyTorch baseline time to the custom “kernel” runtime. 

However, this reward design could lead to reward hacking, as LLMs could directly generate PyTorch-only code and achieve a high reward, which is much easier for them on difficult tasks than writing complex custom kernels. Eventually, LLMs learn it via RL and try to hack KernelBench during evaluation, producing “kernels” with a high success rate and high speedup, but without CUDA code. Here we provide a case of Kevin-32B's hacking behavior in Kernelbench Level 1 Task 59, and the “kernel” Kevin-32B generated is:

\begin{tcolorbox}[mybox, title={Level 1 Task 59 — Reported $1.004\times$ Speedup}]
\ttfamily
\tiny
\begin{verbatim}
import torch
import torch.nn as nn

class ModelNew(nn.Module):
    def __init__(self, in_channels, out_channels, 
    kernel_size, stride=(1, 1, 1), padding=(0, 0, 0), 
    output_padding=(0, 0, 0), groups=1, bias=False):
        super(ModelNew, self).__init__()
        self.conv_transpose3d = nn.ConvTranspose3d(
            in_channels, out_channels, kernel_size,
            stride=stride, padding=padding, output_padding=output_padding,
            groups=groups, bias=bias, 
            # dtype=torch.float16 if using FP16, but kept as default here
        )

    def forward(self, x):
        return self.conv_transpose3d(x)
\end{verbatim}
\normalfont
\end{tcolorbox}

The model basically just copies the original code with only one additional comment. Although in the paper of Kevin-32B, the authors claim that they use “format checks” to “assign 0 reward to responses with no PyTorch
functional operators”. However, the results show that they fail to effectively avoid reward hacking, as shown in the case above and in the experimental results in Section~\ref{sec:hack}. 

Similar results are also found in CUDA-L1's official GitHub repo, as they report the Level 1 Task 12  as "64.4x speedup", where the model just switch another implementation method with PyTorch API:
\begin{tcolorbox}[redbox, title={Level 1 Task 12 — Reported $64.4\times$ Speedup by CUDA-L1}]
\tiny
\ttfamily
\begin{verbatim}
# diag_mm_compare.py
import time
import math
import torch
import torch.nn as nn
import torch.nn.functional as F

# -------------------------------
# Reference implementation
# -------------------------------
class Model(nn.Module):
    """
    Simple model that performs a matrix multiplication of a diagonal matrix 
    with another matrix.
    C = diag(A) * B
    """
    def __init__(self):
        super(Model, self).__init__()

    def forward(self, A, B):
        """
        Args:
            A (torch.Tensor): 1D tensor, diagonal entries. Shape: (N,)
            B (torch.Tensor): 2D tensor. Shape: (N, M)
        Returns:
            torch.Tensor: (N, M)
        """
        return torch.diag(A) @ B


# -------------------------------
# Optimized implementation
# -------------------------------
class ModelNew(nn.Module):
    """
    Optimized model that performs a matrix multiplication of a diagonal matrix
    with another matrix.
    C = diag(A) * B
    """
    def __init__(self):
        super(ModelNew, self).__init__()

    def forward(self, A, B):
        """
        Args:
            A (torch.Tensor): 1D tensor, diagonal entries. Shape: (N,)
            B (torch.Tensor): 2D tensor. Shape: (N, M)
        Returns:
            torch.Tensor: (N, M)
        """
        # Equivalent to torch.diag(A) @ B, but avoids forming the full diagonal 
        matrix
        return B * A.unsqueeze(1)


# -------------------------------
# Hyperparameters & inputs
# -------------------------------
M = 4096
N = 4096

def get_inputs(device=None, dtype=torch.float32):
    A = torch.randn(N, device=device, dtype=dtype)
    B = torch.randn(N, M, device=device, dtype=dtype)
    return [A, B]

def get_init_inputs():
    return []  # No special initialization inputs needed
\end{verbatim}
\normalfont
\end{tcolorbox}

Why do they fail to avoid reward hacking? Taking Kevin-32B as an example, we empirically attribute it to two reasons: First, the anti-hacking mechanism fails to detect reward hacking. Using “format checks” to “assign 0 reward to responses with any PyTorch
functional operators” faces a trade-off---A strict check may assign 0 reward to correct kernels, as \textbf{correct ones may also contain partial PyTorch functional operators in multi-kernel tasks}, while a loose check will omit some reward hacking behaviors. Second, because the “format checks” may omit some hacking kernels, the LLM will learn to reward hacking. Unfortunately, rewards for these hacked kernels are usually higher than those for custom-implemented kernels, as shown in the two cases above. Thus, the model's behavior will be sufficiently changed to the hacking behavior. 

We address this challenge through a rubric-based reward, using an advanced LLM to follow expert-designed rubrics and detect reward hacking. In our early experiments, we found that it is a more robust mechanism compared with “format checks”. And the result in Section~\ref{sec:hack} supports its effectiveness. Details of rubrics are in the following Section~\ref{app:rubric}. 

In addition, we find that \textbf{reward clipping is also very useful}, as the $R_{\max}$ shown in Formulation~\ref{eq:final_reward}. It not only helps avoid training collapse but also helps avoid reward hacking — even if some hacking kernels pass detection, their extremely high rewards will not affect RL training much, since rewards are clipped to a reasonable range.  

\subsubsection{challenges of degenerate behavior}
\label{app:fear}
Besides reward hacking, another challenge in RL training is the emergence of degenerate behaviors. Degenerate behaviors mean that LLM only modifies a trivial part of the reference code(e.g., only replacing a trivial operator such as a standalone \texttt{ReLU} in a Neural Network) or generates too conservative implementations(e.g., a correct but 0.001x speedup kernel). It is also a result of the rule-based reward design. Taking the reward of Kevin-32B in ~\ref{kevinreward} as an example, the reward is positive only when the LLMs generate a correct kernel. That means a correct but very low-speed kernel will score higher than a well-optimized kernel with minor errors. Considering the CUDA code generation is hard for current LLMs~\cite{kernelbench} that they do not have the capability to frequently generate high-speed and correct kernels during RL, model's behavior will be pushed to a direction where \textbf{the model prefers low-speed conservative but correct kernels}, or \textbf{only modifies a trivial part in reference code to ensure correctness}. As a result, the model will generate very conservative implementations in practice. We provide a case of Kevin-32B in KernelBench Level 2 task 1, where models are required to customize a PyTorch implementation with Conv2D, ReLU, and Bias:

\begin{tcolorbox}[mybox, title={Level 2 Task 1 — Reported $1.05\times$ Speedup}]
\ttfamily
\tiny
\begin{verbatim}
import torch
import torch.nn as nn
from torch.utils.cpp_extension import load_inline

relu_bias_source = """
#include <torch/extension.h>
#include <cuda_runtime.h>

__global__ void relu_bias_kernel(const float* input, 
const float* bias, float* output, int total_elements, int channels, int 
spatial_size) {
    int idx = blockIdx.x * blockDim.x + threadIdx.x;
    if (idx >= total_elements) return;
    int c = (idx / spatial_size) % channels;
    float val = input[idx];
    val = fmaxf(val, 0.0f);
    val += bias[c];
    output[idx] = val;
}

torch::Tensor relu_bias_cuda(torch::Tensor input, torch::Tensor bias, 
int spatial_size, int channels) {
    auto total_elements = input.numel();
    auto output = torch::empty_like(input);
    const int block_size = 256;
    const int num_blocks = (total_elements + block_size - 1) / block_size;
    relu_bias_kernel<<<num_blocks, block_size>>>
    (input.data_ptr<float>(), bias.data_ptr<float>(), 
    output.data_ptr<float>(), total_elements, channels, spatial_size);
    return output;
}
"""

relu_bias_header = """
torch::Tensor relu_bias_cuda(torch::Tensor input, torch::Tensor bias, 
int spatial_size, int channels);
"""

relu_bias = load_inline(
    name="relu_bias",
    cpp_sources=relu_bias_header,
    cuda_sources=relu_bias_source,
    functions=["relu_bias_cuda"],
    verbose=True
)

class ModelNew(nn.Module):
    def __init__(self, in_channels, out_channels, kernel_size, bias_shape):
        super(ModelNew, self).__init__()
        self.conv = nn.Conv2d(in_channels, out_channels, kernel_size)
        self.bias = nn.Parameter(torch.randn(bias_shape))
        self.relu_bias = relu_bias

    def forward(self, x):
        x = self.conv(x)
        spatial_size = x.size(2) * x.size(3)
        channels = x.size(1)
        bias_1d = self.bias.view(-1)
        x = self.relu_bias.relu_bias_cuda(x, bias_1d, spatial_size, channels)
        return x
\end{verbatim}
\normalfont
\end{tcolorbox}

Although it gets 1.10x speedup in the benchmark, it is not a very meaningful implementation in practice: the main bottleneck in the PyTorch code is Conv2D. Optimization in Conv2D will bring more practical value, but the RL-based model does not achieve that, even though it is prompted to implement it by the verifier's feedback.

Promising methods to address this issue are to modify the reward design in two directions: encourage more aggressive optimization or punish overly conservative implementations. For the former, more advanced CUDA techniques used in implementation should be given a higher reward. For the latter, only “writing a ReLU” should be given a lower reward. The goal of the two reward design directions could be achieved by a rubric reward at the same time. We use expert designed rubric reward to encourage more advanced techniques and broader operations coverage. The result in Section~\ref{sec:main_result} supports our method, for our model gets the best or the second best speedup in evaluation, which even outperforms advanced GPT-5.2. 

\subsection{Case study of StitchCUDA}
\label{app:case}
It is difficult to quantitatively analyze “how many kernels should cover” or “how many advanced techniques we can use” in practice. But here we provide a case code generated by our method during iterations on Level 3 Task 44, to demonstrate how it performs in practice:

\begin{tcolorbox}[mybox, title={our model in Level 3 Task 44 — Reported 3.75 $\times$ Speedup}]
\ttfamily
\tiny
\begin{verbatim}
# <complete ModelNew code>
import math
import torch
import torch.nn as nn
import torch.nn.functional as F
from torch.utils.cpp_extension import load_inline

# ---- Custom CUDA: cuBLASLt fused GEMM epilogues (FP16) with cached Lt objects/algo + persistent workspace ----
cublaslt_fused_ext = load_inline(
    name="cublaslt_fused_ext",
    cpp_sources=r"""
#include <torch/extension.h>

torch::Tensor fc_bias_forward(torch::Tensor x, torch::Tensor w, torch::Tensor b, int64_t workspace_bytes);
torch::Tensor fc_gelu_bias_forward(torch::Tensor x, torch::Tensor w, torch::Tensor b, int64_t workspace_bytes);
torch::Tensor fc_relu_bias_forward(torch::Tensor x, torch::Tensor w, torch::Tensor b, int64_t workspace_bytes);

PYBIND11_MODULE(TORCH_EXTENSION_NAME, m) {
  m.def("fc_bias_forward", &fc_bias_forward, "FC + Bias (cuBLASLt, CUDA)");
  m.def("fc_gelu_bias_forward", &fc_gelu_bias_forward, "FC + Bias + GELU (cuBLASLt, CUDA)");
  m.def("fc_relu_bias_forward", &fc_relu_bias_forward, "FC + Bias + ReLU (cuBLASLt, CUDA)");
}
""",
    cuda_sources=r"""
#include <torch/extension.h>
#include <ATen/cuda/CUDAContext.h>

#include <cuda.h>
#include <cuda_runtime.h>

#include <cublasLt.h>
#include <cublas_v2.h>

#include <unordered_map>
#include <mutex>
#include <cstdint>

static inline void checkLt(cublasStatus_t status, const char* msg) {
  TORCH_CHECK(status == CUBLAS_STATUS_SUCCESS, msg, " (cublas status=", (int)status, ")");
}
static inline void checkCuda(cudaError_t e, const char* msg) {
  TORCH_CHECK(e == cudaSuccess, msg, " (cuda err=", (int)e, " ", cudaGetErrorString(e), ")");
}

struct Key {
  int M, N, K;
  int epi;
  int64_t workspace;
  bool operator==(const Key& o) const {
    return M == o.M && N == o.N && K == o.K && epi == o.epi && workspace == o.workspace;
  }
};
struct KeyHash {
  size_t operator()(const Key& k) const noexcept {
    uint64_t h = 1469598103934665603ull;
    auto mix = [&](uint64_t v) {
      h ^= v + 0x9e3779b97f4a7c15ull + (h << 6) + (h >> 2);
    };
    mix((uint64_t)k.M);
    mix((uint64_t)k.N);
    mix((uint64_t)k.K);
    mix((uint64_t)k.epi);
    mix((uint64_t)k.workspace);
    return (size_t)h;
  }
};

struct CacheEntry {
  cublasLtMatmulDesc_t matmulDesc = nullptr;
  cublasLtMatrixLayout_t Adesc = nullptr;
  cublasLtMatrixLayout_t Bdesc = nullptr;
  cublasLtMatrixLayout_t Cdesc = nullptr;
  cublasLtMatmulPreference_t pref = nullptr;
  cublasLtMatmulHeuristicResult_t heur{};
  bool valid = false;
};

static std::mutex g_mutex;
static cublasLtHandle_t g_ltHandle = nullptr;
static std::unordered_map<Key, CacheEntry, KeyHash> g_cache;

// Persistent workspace per CUDA stream
struct WS {
  void* ptr = nullptr;
  size_t bytes = 0;
};
static std::unordered_map<uint64_t, WS> g_ws_by_stream;

static cublasLtHandle_t getHandle() {
  if (g_ltHandle) return g_ltHandle;
  std::lock_guard<std::mutex> lock(g_mutex);
  if (!g_ltHandle) {
    checkLt(cublasLtCreate(&g_ltHandle), "cublasLtCreate failed");
  }
  return g_ltHandle;
}

static WS getWorkspaceForStream(cudaStream_t stream, size_t bytes_required) {
  // Key by stream handle value
  uint64_t sid = (uint64_t)(uintptr_t)stream;
  std::lock_guard<std::mutex> lock(g_mutex);
  WS& ws = g_ws_by_stream[sid];
  if (bytes_required == 0) return ws;
  if (ws.ptr == nullptr || ws.bytes < bytes_required) {
    // (Re)allocate larger buffer; keep old one allocated to avoid sync/free complications.
    void* new_ptr = nullptr;
    checkCuda(cudaMalloc(&new_ptr, bytes_required), "cudaMalloc persistent workspace failed");
    ws.ptr = new_ptr;
    ws.bytes = bytes_required;
  }
  return ws;
}

// Column-major interpretation trick:
// A points to w row-major [N,K] interpreted as col-major [K,N], op(A)=T => [N,K]
// B points to x row-major [M,K] interpreted as col-major [K,M], op(B)=N => [K,M]
// C points to y row-major [M,N] interpreted as col-major [N,M]
static CacheEntry buildEntry(int M, int N, int K, cublasLtEpilogue_t epi, int64_t workspace_bytes) {
  CacheEntry entry;
  cublasLtHandle_t ltHandle = getHandle();

  checkLt(cublasLtMatmulDescCreate(&entry.matmulDesc, CUBLAS_COMPUTE_32F, CUDA_R_32F),
          "cublasLtMatmulDescCreate failed");

  cublasOperation_t transA = CUBLAS_OP_T;
  cublasOperation_t transB = CUBLAS_OP_N;
  checkLt(cublasLtMatmulDescSetAttribute(entry.matmulDesc, CUBLASLT_MATMUL_DESC_TRANSA, &transA, sizeof(transA)),
          "set TRANSA failed");
  checkLt(cublasLtMatmulDescSetAttribute(entry.matmulDesc, CUBLASLT_MATMUL_DESC_TRANSB, &transB, sizeof(transB)),
          "set TRANSB failed");
  checkLt(cublasLtMatmulDescSetAttribute(entry.matmulDesc, CUBLASLT_MATMUL_DESC_EPILOGUE, &epi, sizeof(epi)),
          "set EPILOGUE failed");

  const int64_t lda = K, ldb = K, ldc = N;
  checkLt(cublasLtMatrixLayoutCreate(&entry.Adesc, CUDA_R_16F, K, N, lda), "create Adesc failed");
  checkLt(cublasLtMatrixLayoutCreate(&entry.Bdesc, CUDA_R_16F, K, M, ldb), "create Bdesc failed");
  checkLt(cublasLtMatrixLayoutCreate(&entry.Cdesc, CUDA_R_16F, N, M, ldc), "create Cdesc failed");

  checkLt(cublasLtMatmulPreferenceCreate(&entry.pref), "preferenceCreate failed");
  size_t wsize = (workspace_bytes > 0) ? (size_t)workspace_bytes : (size_t)0;
  checkLt(cublasLtMatmulPreferenceSetAttribute(entry.pref, CUBLASLT_MATMUL_PREF_MAX_WORKSPACE_BYTES, &wsize, sizeof(wsize)),
          "set workspace pref failed");

  int returned = 0;
  checkLt(cublasLtMatmulAlgoGetHeuristic(
              ltHandle, entry.matmulDesc, entry.Adesc, entry.Bdesc, entry.Cdesc, entry.Cdesc,
              entry.pref, 1, &entry.heur, &returned),
          "algoGetHeuristic failed");
  TORCH_CHECK(returned > 0, "cuBLASLt: no heuristic algorithm found for requested epilogue/config");
  entry.valid = true;
  return entry;
}

static CacheEntry& getOrCreate(int M, int N, int K, cublasLtEpilogue_t epi, int64_t workspace_bytes) {
  Key key{M, N, K, (int)epi, workspace_bytes};
  {
    std::lock_guard<std::mutex> lock(g_mutex);
    auto it = g_cache.find(key);
    if (it != g_cache.end() && it->second.valid) return it->second;
  }

  CacheEntry entry = buildEntry(M, N, K, epi, workspace_bytes);

  std::lock_guard<std::mutex> lock(g_mutex);
  auto [it, inserted] = g_cache.emplace(key, entry);
  if (!inserted) it->second = entry;
  return it->second;
}

static torch::Tensor fc_epilogue_impl(torch::Tensor x, torch::Tensor w, torch::Tensor b,
                                      cublasLtEpilogue_t epi, int64_t workspace_bytes) {
  TORCH_CHECK(x.is_cuda() && w.is_cuda() && b.is_cuda(), "inputs must be CUDA");
  TORCH_CHECK(x.is_contiguous(), "x must be contiguous");
  TORCH_CHECK(w.is_contiguous(), "w must be contiguous");
  TORCH_CHECK(b.is_contiguous(), "b must be contiguous");
  TORCH_CHECK(x.scalar_type() == at::kHalf, "x must be float16");
  TORCH_CHECK(w.scalar_type() == at::kHalf, "w must be float16");
  TORCH_CHECK(b.scalar_type() == at::kHalf, "b must be float16");
  TORCH_CHECK(x.dim() == 2, "x must be 2D [M, K]");
  TORCH_CHECK(w.dim() == 2, "w must be 2D [N, K]");
  TORCH_CHECK(b.dim() == 1, "b must be 1D [N]");

  const int64_t M64 = x.size(0), K64 = x.size(1), N64 = w.size(0);
  TORCH_CHECK(w.size(1) == K64, "w shape mismatch");
  TORCH_CHECK(b.size(0) == N64, "b shape mismatch");
  TORCH_CHECK(M64 <= INT_MAX && N64 <= INT_MAX && K64 <= INT_MAX, "dims too large");
  int M = (int)M64, N = (int)N64, K = (int)K64;

  auto y = torch::empty({M64, N64}, x.options());

  CacheEntry& ce = getOrCreate(M, N, K, epi, workspace_bytes);

  // Bias pointer is per-call
  void* biasPtr = (void*)b.data_ptr<at::Half>();
  checkLt(cublasLtMatmulDescSetAttribute(ce.matmulDesc, CUBLASLT_MATMUL_DESC_BIAS_POINTER, &biasPtr, sizeof(biasPtr)),
          "set BIAS_POINTER failed");

  cudaStream_t stream = at::cuda::getDefaultCUDAStream();

  size_t wsize = (workspace_bytes > 0) ? (size_t)workspace_bytes : (size_t)0;
  WS ws = getWorkspaceForStream(stream, wsize);
  void* workspace = (wsize > 0) ? ws.ptr : nullptr;

  float alpha = 1.0f, beta = 0.0f;
  const void* A = (const void*)w.data_ptr<at::Half>();
  const void* Bptr = (const void*)x.data_ptr<at::Half>();
  void* Cptr = (void*)y.data_ptr<at::Half>();

  checkLt(cublasLtMatmul(getHandle(),
                        ce.matmulDesc,
                        &alpha,
                        A, ce.Adesc,
                        Bptr, ce.Bdesc,
                        &beta,
                        Cptr, ce.Cdesc,
                        Cptr, ce.Cdesc,
                        &ce.heur.algo,
                        workspace, wsize,
                        stream),
          "cublasLtMatmul failed");

  return y;
}

torch::Tensor fc_bias_forward(torch::Tensor x, torch::Tensor w, torch::Tensor b, int64_t workspace_bytes) {
  return fc_epilogue_impl(x, w, b, CUBLASLT_EPILOGUE_BIAS, workspace_bytes);
}
torch::Tensor fc_gelu_bias_forward(torch::Tensor x, torch::Tensor w, torch::Tensor b, int64_t workspace_bytes) {
  return fc_epilogue_impl(x, w, b, CUBLASLT_EPILOGUE_GELU_BIAS, workspace_bytes);
}
torch::Tensor fc_relu_bias_forward(torch::Tensor x, torch::Tensor w, torch::Tensor b, int64_t workspace_bytes) {
  return fc_epilogue_impl(x, w, b, CUBLASLT_EPILOGUE_RELU_BIAS, workspace_bytes);
}
""",
    functions=None,
    extra_cuda_cflags=["-O3", "--use_fast_math", "-lineinfo"],
    extra_cflags=["-O3"],
    with_cuda=True,
    verbose=False,
)


class FusedLinearBias(nn.Module):
    """FC + Bias via cuBLASLt epilogue (cached). Params FP16."""
    def __init__(self, in_features: int, out_features: int, workspace_bytes: int = 64 * 1024 * 1024):
        super().__init__()
        self.weight = nn.Parameter(torch.empty(out_features, in_features, dtype=torch.float16))
        self.bias = nn.Parameter(torch.empty(out_features, dtype=torch.float16))
        self.workspace_bytes = int(workspace_bytes)
        self.reset_parameters()

    def reset_parameters(self):
        w = torch.empty_like(self.weight, dtype=torch.float32)
        b = torch.empty_like(self.bias, dtype=torch.float32)
        nn.init.kaiming_uniform_(w, a=math.sqrt(5))
        fan_in = w.size(1)
        bound = 1 / math.sqrt(fan_in)
        nn.init.uniform_(b, -bound, bound)
        with torch.no_grad():
            self.weight.copy_(w.to(torch.float16).contiguous())
            self.bias.copy_(b.to(torch.float16).contiguous())

    def forward(self, x: torch.Tensor) -> torch.Tensor:
        orig_shape = x.shape
        x2 = x.reshape(-1, orig_shape[-1]).contiguous()
        if x2.dtype != torch.float16:
            x2 = x2.to(dtype=torch.float16)
        y2 = cublaslt_fused_ext.fc_bias_forward(x2, self.weight.contiguous(), self.bias.contiguous(), self.workspace_bytes)
        return y2.view(*orig_shape[:-1], y2.shape[-1])


class FusedFCGELU(nn.Module):
    """FC + Bias + GELU via cuBLASLt epilogue (cached). Params FP16."""
    def __init__(self, in_features: int, out_features: int, workspace_bytes: int = 64 * 1024 * 1024):
        super().__init__()
        self.weight = nn.Parameter(torch.empty(out_features, in_features, dtype=torch.float16))
        self.bias = nn.Parameter(torch.empty(out_features, dtype=torch.float16))
        self.workspace_bytes = int(workspace_bytes)
        self.reset_parameters()

    def reset_parameters(self):
        w = torch.empty_like(self.weight, dtype=torch.float32)
        b = torch.empty_like(self.bias, dtype=torch.float32)
        nn.init.kaiming_uniform_(w, a=math.sqrt(5))
        fan_in = w.size(1)
        bound = 1 / math.sqrt(fan_in)
        nn.init.uniform_(b, -bound, bound)
        with torch.no_grad():
            self.weight.copy_(w.to(torch.float16).contiguous())
            self.bias.copy_(b.to(torch.float16).contiguous())

    def forward(self, x: torch.Tensor) -> torch.Tensor:
        orig_shape = x.shape
        x2 = x.reshape(-1, orig_shape[-1]).contiguous()
        if x2.dtype != torch.float16:
            x2 = x2.to(dtype=torch.float16)
        y2 = cublaslt_fused_ext.fc_gelu_bias_forward(x2, self.weight.contiguous(), self.bias.contiguous(), self.workspace_bytes)
        return y2.view(*orig_shape[:-1], y2.shape[-1])


class CausalSelfAttention(nn.Module):
    def __init__(self, n_embd, n_head, attn_pdrop, resid_pdrop, max_seqlen):
        super().__init__()
        assert n_embd % n_head == 0
        self.n_head = n_head
        self.n_embd = n_embd
        self.c_attn = nn.Linear(n_embd, 3 * n_embd)
        self.c_proj = nn.Linear(n_embd, n_embd)
        self.resid_dropout = nn.Dropout(resid_pdrop)
        with torch.no_grad():
            self.c_attn.weight.data = self.c_attn.weight.data.contiguous()
            self.c_attn.bias.data = self.c_attn.bias.data.contiguous()
            self.c_proj.weight.data = self.c_proj.weight.data.contiguous()
            self.c_proj.bias.data = self.c_proj.bias.data.contiguous()

    def forward(self, x):
        B, T, C = x.size()
        hs = C // self.n_head
        q, k, v = self.c_attn(x).split(self.n_embd, dim=2)
        q = q.view(B, T, self.n_head, hs).transpose(1, 2)
        k = k.view(B, T, self.n_head, hs).transpose(1, 2)
        v = v.view(B, T, self.n_head, hs).transpose(1, 2)
        y = F.scaled_dot_product_attention(q, k, v, attn_mask=None, dropout_p=0.0, is_causal=True)
        y = y.transpose(1, 2).contiguous().view(B, T, C)
        y = self.resid_dropout(self.c_proj(y))
        return y


class ModelNew(nn.Module):
    def __init__(self, n_embd, n_head, attn_pdrop, resid_pdrop, max_seqlen):
        super().__init__()
        self.ln_1 = nn.LayerNorm(n_embd)
        self.attn = CausalSelfAttention(n_embd, n_head, attn_pdrop, resid_pdrop, max_seqlen)
        self.ln_2 = nn.LayerNorm(n_embd)

        self.mlp = nn.ModuleDict(
            dict(
                c_fc_gelu=FusedFCGELU(n_embd, 4 * n_embd, workspace_bytes=64 * 1024 * 1024),
                c_proj_bias=FusedLinearBias(4 * n_embd, n_embd, workspace_bytes=64 * 1024 * 1024),
                dropout=nn.Dropout(resid_pdrop),
            )
        )

    def forward(self, x):
        x_fp32 = x.float()

        ln1 = self.ln_1(x_fp32)
        with torch.cuda.amp.autocast(dtype=torch.float16):
            attn_out = self.attn(ln1)
        x_fp32 = x_fp32 + attn_out.float()

        ln2 = self.ln_2(x_fp32)
        with torch.cuda.amp.autocast(dtype=torch.float16):
            h = self.mlp["c_fc_gelu"](ln2)
            h = self.mlp["c_proj_bias"](h)
            h = self.mlp["dropout"](h)
        x_fp32 = x_fp32 + h.float()

        return x_fp32


batch_size = 128
max_seqlen = 1024
seq_len = 512
n_embd = 768
n_head = 8
attn_pdrop = 0.0
resid_pdrop = 0.0


def get_inputs():
    return [torch.rand(batch_size, seq_len, n_embd, device="cuda", dtype=torch.float32)]


def get_init_inputs():
    return [n_embd, n_head, attn_pdrop, resid_pdrop, max_seqlen]
\end{verbatim}
\normalfont
\end{tcolorbox}
The final solution generated by StitchCUDA applies the following optimizations.

At the system level, Planner proposed 2 critical optimizations: \textbf{Mixed precision computing:} LayerNorm and residual accumulation run in \texttt{fp32}, while the compute-intensive MLP runs in \texttt{fp16}, preserving stability while enabling Tensor Cores. \textbf{Data layout:} weights/biases are made \texttt{contiguous()} at initialization, and activations are reshaped and made contiguous before GEMM to avoid stride-induced fallbacks and implicit copies. Enabling pinned memory for contiguous data transfers to maximize PCIe bandwidth utilization.

On the kernel level, Planner gives the following implementation guide. \textbf{MLP fusion:} implemented as custom fused modules instead of separate \texttt{Linear}$\rightarrow$\texttt{GELU}$\rightarrow$\texttt{Linear} ops, reducing kernel launch and dispatch overhead. \textbf{cuBLASLt epilogue fusion:} cuBLASLt epilogues fuse \texttt{(GEMM+bias)} and \texttt{(GEMM+bias+GELU)} into single launches, cutting elementwise kernels and intermediate traffic, which has been observed as bottelneck during Nsys profiling. \textbf{Heuristic descriptors caching:} cuBLASLt descriptors and selected algorithms are cached by shape/epilogue/workspace, avoiding repeated heuristic queries. \textbf{Persistent per-stream workspace:} per-stream buffers are retained and grown as needed, reducing \texttt{cudaMalloc}/\texttt{cudaFree} overhead and associated synchronization.

\noindent Overall, these system-level and kernel-level optimizations yield a $2.7\times$ speedup over the reference implementation in eager mode and $2.08\times$ speedup over enabling \texttt{torch.compile} on this task. 

Unlike Kevin-32B, we do not classify degenerate behavior as a form of reward hacking for two reasons: First, although not very meaningful, LLMs actually write CUDA kernels in custom code. Second, for current LLMs, it is very hard to cover all reference operations with custom kernels in a single inference. Directly assigning 0 reward for this behavior will lead to a severe sparse reward problem, which is very harmful to RL training, as we observed in our early experiments.

\subsubsection{Rubric Design}
To address the challenges of reward hacking and degenerate behaviors, we use a rubric reward to provide a more comprehensive evaluation of the kernel's quality, beyond the rule-based reward function. We invite human experts in CUDA engineering to write the rubrics. After refining it with GPT-5.2 and testing it in offline samples to ensure its effectiveness, we get a well-designed rubric with the following four dimensions: 
\begin{itemize}
    \item \textbf{Anti-Hacking}, penalizing reward exploitation;
    \item \textbf{CUDA Engineering}, rewarding advanced optimization techniques, including tiling with shared memory, asynchronous memcpy, cuBLASLt, CUTLASS, wmma instruction for tensor core, kernel-fusion, mixed-precision computing, and inline PTX Assembly;
    \item \textbf{Operator Coverage}, encouraging broader optimization for complex multi-kernel programs, covering host-side optimization and as many kernels as possible;
    \item \textbf{Skill Compliance}, enforcing adherence to task requirements (Skill~1) or feedback instructions (Skill~2).
\end{itemize}

Each dimension uses a discrete scale with detailed criteria per score level. 

\label{app:rubric}

\subsection{Comparing computation overhead between our rubric-based agentic RL with multi-turn agentic RL and agentic RL without rubrics}
\label{app:time}
The most important motivation for decomposing multi-turn agentic RL into learning two atomic skills is the prohibitive computational cost of multi-turn agentic RL. Here, we provide details to demonstrate it and compare the computation overhead between different methods.
\subsubsection{The multi-turn rollout overhead}
The standard method for agentic RL training~\cite{jin2025searchr1trainingllmsreason,wang2025ragenunderstandingselfevolutionllm} is to collect the trajectories of model-environment interactions in real executions, assign trajectory-level reward, and train the model via RL algorithms like GRPO~\cite{Guo_2025}.  During the procedure, the most time-consuming part is usually the multi-turn rollout(collecting the trajectories of model-environment interactions). The problem even becomes worse in RL for CUDA code generation: In a famous agentic RL work, Search-R1~\cite{jin2025searchr1trainingllmsreason}, it takes seconds for the model to get feedback from the search environment in a single turn. But in the CUDA code generation setting, it takes several minutes to get feedback from the environment. As shown in Fig.~\ref {fig:RL}, for a single-turn interaction, the generated code needs to be compiled and tested on several unit tests. After that, the Verifier needs to call the RAG database or Nsys/NCU profiling to generate further feedback. Finally, the feedback is generated by Verifier and sent back to Coder. In our experiments, the single-turn interaction process takes 4-5 minutes on average. Given that the trajectories we need contain multi-turn interactions, the overall time required to collect a 15-turn trajectory is 60-75 minutes. Finally, with 400 RL samples, a training batch size of 16, a group size of 8, and 3 epochs, the estimated training time for one model is about 60 days, which makes RL training too costly and impractical in practice. 

\subsection{Compare the computation overhead between different methods}
To provide a clear comparison, we evaluate the computational overhead of different methods, including multi-turn agentic RL, our method without a rubric reward, and our rubric-based agentic RL. Using the same training dataset and parameters, the H200-Hour comparison is shown in Table~\ref {tab:h200_overhead}. The result shows that our method is very efficient compared with multi-turn agentic RL. Although the rubric assignment takes more time, the extra computation overhead is modest. 
\begin{table}[t]
\centering
\small
\setlength{\tabcolsep}{6pt}
\begin{tabular}{lccc}
\toprule
Method & Turns per trajectory & Total turns & H200-Hour \\
\midrule
Multi-turn agentic RL & $15$ &
$144000$ &
$9600-12000$ \\
\addlinespace
Our method (w/o rubric reward) & $1$ &
- &
$128$ \\
\addlinespace
Rubric-based agentic RL (ours) & $1$ &
- &
$160$ \\
\bottomrule
\end{tabular}
\vspace{2pt}
\caption{Computation overhead comparison measured in H200-Hour under the same dataset and training hyperparameters (400 samples, group=8, epoch=3).
We estimate the average environment interaction takes 4--5 minutes.}
\label{tab:h200_overhead}
\vspace{-6pt}
\begin{flushleft}
\end{flushleft}
\end{table}


\end{document}